\documentclass[fleqn,usenatbib]{mnras}

\usepackage[T1]{fontenc}
\usepackage{ae,aecompl}
\usepackage{graphicx}         
\usepackage{amsmath}          
\usepackage{amssymb}          
\usepackage{array}
\usepackage{float}
\usepackage{etoolbox}
\usepackage{CJK}              
\usepackage{newtxtext,newtxmath}

\newcommand {\eV}             {\,\rm eV}
\newcommand {\pc}             {\,\rm pc}
\newcommand {\kpc}            {\,\rm kpc}
\newcommand {\Mpc}            {\,\rm Mpc}
\newcommand {\pch}            {\,{h^{-1}\pc}}
\newcommand {\kpch}           {\,{h^{-1}\kpc}}
\newcommand {\Mpch}           {\,{h^{-1}\Mpc}}
\newcommand {\MpchInv}        {\,{h\Mpc^{-1}}}
\newcommand {\Msun}           {\,\rm{M}_{\sun}}

\newcommand {\gamer}          {\textsc{gamer-2}}
\newcommand {\psiDM}          {\psi \rm{DM}}
\newcommand {\mM}             {m_{\rm major}}
\newcommand {\mm}             {m_{\rm minor}}
\newcommand {\mratio}         {\mM:\mm}
\newcommand {\Mh}             {M_{\rm halo}}
\newcommand {\rcM}            {r_{\rm c, major}}
\newcommand {\rcm}            {r_{\rm c, minor}}
\newcommand {\McM}            {M_{\rm c, major}}
\newcommand {\Mcm}            {M_{\rm c, minor}}
\newcommand {\dmaxM}          {\rho_{\rm max, major}}
\newcommand {\dmaxm}          {\rho_{\rm max, minor}}

\newcommand {\sref}[1]        {Section~\ref{#1}}
\newcommand {\fref}[1]        {Fig.~\ref{#1}}

\newcommand {\eref}[1]        {Eq.~(\ref{#1})}

\newcommand {\be}             {\begin{equation}}
\newcommand {\ee}             {\end{equation}}

\newcommand{\vect}[1]         {\boldsymbol{#1}}

\begin{document}
\begin{CJK*}{UTF8}{bkai}

\title[Two-mass $\psiDM$]{Cosmological Simulations of Two-Component Wave Dark Matter}

\author[H. Huang et al.]{
Hsinhao Huang (黃新豪)$^{1}$,
Hsi-Yu Schive (薛熙于)$^{1,2,3,4}$\thanks{E-mail: hyschive@phys.ntu.edu.tw},
Tzihong Chiueh (闕志鴻)$^{1,2,3}$
\vspace*{8pt}
\\
$^{1}$Department of Physics, National Taiwan University, Taipei 10617, Taiwan\\
$^{2}$Institute of Astrophysics, National Taiwan University, Taipei 10617, Taiwan\\
$^{3}$Center for Theoretical Physics, National Taiwan University, Taipei 10617, Taiwan\\
$^{4}$Physics Division, National Center for Theoretical Sciences, Taipei 10617, Taiwan}

\date{Accepted XXX. Received YYY; in original form ZZZ}

\pubyear{2023}
\label{firstpage}
\pagerange{\pageref{firstpage}--\pageref{lastpage}}
\maketitle
\end{CJK*}

\begin{abstract}
Wave (fuzzy) dark matter ($\psiDM$) consists of ultralight bosons,
featuring a solitonic core within a granular halo.
Here we extend $\psiDM$ to two components, with distinct
particle masses $m$ and coupled only through gravity, and investigate the
resulting soliton-halo structure via cosmological simulations.
Specifically, we assume $\psiDM$ contains $75$ per cent major component
and $25$ per cent minor component, fix the major-component particle mass to
$\mM=1\times10^{-22}\eV$, and explore two different minor-component
particle masses with $\mratio=3:1$ and $1:3$, respectively.
For $\mratio=3:1$, we find that
(i) the major- and minor-component solitons
coexist, have comparable masses, and are roughly concentric.
(ii) The soliton peak density is significantly lower than the single-component
counterpart, leading to a smoother soliton-to-halo transition and
rotation curve.
(iii) The combined soliton mass of both components follows the same
single-component core-halo mass relation.
In dramatic contrast, for $\mratio=1:3$, a minor-component
soliton cannot form with the presence of a stable major-component soliton;
the total density profile, for both halo and soliton, is thus dominated by the
major component and closely follows the single-component case.
To support this finding, we propose a toy model illustrating that it is
difficult to form a soliton in a hot environment associated with a deep
gravitational potential.
The work demonstrates the extra flexibility added to the multi-component $\psiDM$
model can resolve observational tensions over the
single-component model while retaining its key features.
\end{abstract}


\section{Introduction}
\label{sec:intro}
There is increasing astrophysical evidence for the existence of dark matter (DM),
which interacts primarily through gravity and is likely beyond the
Standard Model of particle physics.
See \citet{Bertone2018RvMP} for a review of the history of DM discoveries.
The nature of DM remains a mystery \citep{Feng2010ARA&A, Arun2017AdSpR}.
For example, the particle masses of different DM candidates span
many orders of magnitude, leading to a large parameter space to be searched.
To date, direct detection experiments have not found any promising DM candidate
\citep[e.g.][]{Bernabei2013, FermiLAT2015JCAP, Angloher2016EPJC,
Cui2017PhRvL, Liu2017NatPh}.

Cold dark matter (CDM) consists of non-relativistic collisionless particles,
which, together with a cosmological constant $\Lambda$, can successfully
explain the large-scale structure of the universe and the
Cosmic Microwave Background \citep{Planck2020A&A, Bull2016PDU}.
However, some of the CDM predictions from numerical simulations are not fully
consistent with observations on small scales \citep{Weinberg2015PNAS}.
These small-scale problems include, for instance,
the cusp-core problem of halo density profiles \citep{Navarro1996ApJ, Navarro1997ApJ, deBlok2010AdAst},
the number of observed satellite galaxies being fewer than expected
\citep{Moore1999ApJ, Klypin1999ApJ},
the too-big-to-fail problem of subhaloes \citep{Boylan-Kolchin2011MNRAS},
and the diversity problem \citep{Bullock2017ARA&A}.
Baryonic physics may provide solutions to these challenges
\citep{Chan2015MNRAS, Sawala2016MNRAS, DelPopolo2018PhRvD, Read2019MNRAS, Sales2022}.

Wave dark matter \citep{Hui2021ARA&A} is one of the emerging
alternative models of CDM to solve the small-scale problems.
In this scenario, DM is made up of light bosons with a very large occupation
number such that it is best described by a macroscopic wave function
obeying the Schr\"{o}dinger equation.
Especially, for an ultralight DM particle mass
($m \sim 10^{-22}\textrm{--}10^{-20}\eV$), the model is generally referred
to as fuzzy dark matter \citep[FDM or $\psiDM$;][]{Hu2000PhRvL}, with a
de Broglie wavelength on the galactic scale.
Axions or axion-like particles predicted by string theory are promising candidates
for $\psiDM$ \citep{Marsh2016PhR, ChadhaDay2022SciA},
which can be produced in a cosmological context \citep{Arvanitaki2020PhRvD}.
For recent reviews of $\psiDM$, see
\citet{Hui2017PhRvD, UrenaLopez2019FrASS, Niemeyer2020PrPNP, Hui2021ARA&A, Ferreira2021A&ARv}.

While the large-scale structure of $\psiDM$ is indistinguishable from CDM,
$\psiDM$ predicts the suppression of small-scale structure as a result of
quantum pressure, thus providing a plausible solution to the CDM small-scale
problems \citep[e.g.][]{Matos2001PhRvD, Schive2014a, Marsh2015MNRAS, Chen2017}.
The cosmological structure formation in the $\psiDM$ scenario has been
intensively studied recently
\citep[e.g.][]{Woo2009ApJ, Schive2014a, Schive2016ApJ, Veltmaat2016PhRvD,
Du2017MNRAS, Veltmaat2018PhRvD, Zhang2018FrASS, Zhang2018ApJ, Nori2018MNRAS,
Nori2019MNRAS, Li2019PhRvD, Mina2022A&A, Mocz2020MNRAS, Nori2021MNRAS, May2021MNRAS}.
One of the most distinctive features coming from its wave nature is the
solitonic core forming at the centre of every dark matter halo \citep{Schive2014a}.
Recent investigations on solitons include
their formation and interaction \citep[e.g.][]{Schwabe2016PhRvD, Amin2019PhRvD, Hertzberg2020JCAP},
core-halo relation \citep[e.g.][]{Schive2014PhRvL, Mocz2017MNRAS, Nori2021MNRAS, Chan2022MNRAS},
soliton random motion \citep{Schive2020, Li2021, Chiang2021, Chowdhury2021},
and their astrophysical impacts \citep[e.g.][]{Li2020, Bar2022}.

For spin-0 bosons, $\psiDM$ can be described by a scalar field, also known as
scalar DM. Relatedly, vector DM (or dark photon DM) considers spin-1 bosons
\citep[e.g.][]{Dimopoulos2006PhRvD, Nelson2011PhRvD, Arias2012JCAP, Cembranos2017JHEP,
Baryakhtar2017PhRvD, Adshead2021PhRvD, Blinov2021PhRvD, Gorghetto2022JCAP}.
Numerical simulations of vector DM show that the halo density profile
is smoother compared to scalar DM \citep{Amin2022JCAP}.
For vector DM, the dynamical equations of motion are the same as three equal-mass
scalar fields coupled through gravity and possible
spin-spin self-interactions \citep{Jain2022arXiv}. In contrast, in this work we
investigate DM consisting of two scalar fields with distinct particle masses.
Fluctuations of two-component scalar field of different particle masses have
recently also been investigated, which may have some connections to
the earliest galaxies found by \textit{JWST} \citep{Hsu2021PhRvD,CurtisLake2023NatAs}.

There are several motivations for considering $\psiDM$ with multiple particle masses.
First, the particle mass constraints from different observations, such as
dwarf galaxies, Lyman-$\alpha$ forest, galactic rotation curves, stellar streams
and black hole spins
\citep{Li2014PhRvD, Calabrese2016MNRAS, GonzalezMorales2017MNRAS, Armengaud2017MNRAS,
Kobayashi2017PhRvD, Irsic2017PhRvL, Schutz2020PhRvD, Benito2020PhRvD, Rogers2021PhRvL,
Nadler2021PhRvL, Banik2021JCAP, Chan2021ApJ, Bar2022, Dalal2022PhRvD, Unal2021JCAP},
are not fully consistent with each other.
Similarly, the soliton profiles with a single fixed particle mass have difficulties
matching different observational constraints
\citep[e.g.][]{Deng2018PhRvD, Bar2018PhRvD, Burkert2020ApJ, Safarzadeh2020ApJ, Kendall2020PASA}.
It is therefore important to investigate whether a multi-component $\psiDM$
scenario could provide a solution to these challenges.

Second, multiple components of $\psiDM$ can arise naturally from particle physics.
For example, axions (or axion-like particles) with a mass spectrum are well motivated
in string theory
\citep[the so-called `String Axiverse';][]{Svrcek2006JHEP, Arvanitaki2010PhRvD, Cicoli2012JHEP, Cicoli2022JHEP}
and clockwork axions \citep{Kaplan2016PhRvD}. These axions are expected to exist
simultaneously and all act like $\psiDM$ candidates.

Third, there is no genuine multi-component $\psiDM$ simulation so far\footnote{
After submitting this paper, there are some three-dimensional simulations to
investigate the dynamics and properties of
constructed haloes \citep{Gosenca2023PhRvD}
and solitons \citep{Glennon2023PhRvD}
in a multi-component ultralight dark matter model.}.
Although some studies already consider a multi-component model
\citep[e.g.][]{Kan2017PhRvD, Eby2020JCAP, Luu2020, Berman2020, Guo2021JCAP,
Street2022PhRvD, Cyncynates2022PhRvD, TellezTovar2022PhRvD},
they are based on either analytical approaches or toy models with
spherical symmetry.
Therefore, three-dimensional cosmological simulations
are indispensable for scrutinizing this scenario further.

Driven by these motivations, in this proof-of-concept study we assume
$\psiDM$ is composed of two components with distinct particle masses, coupled
only through gravity, and conduct cosmological simulations to investigate
the resulting soliton-halo structure. We also explore the results with
different particle mass ratios between the two components.
The paper is organized as follows.
In \sref{sec:numer}, we describe the governing equations and simulation set-up.
The simulation results are presented in \sref{sec:resul}.
We then discuss various aspects of our findings in \sref{sec:discu} and conclude
in \sref{sec:concl}. \sref{app:code_verify} provides the verification
of the accuracy of our simulation code.

\section{Numerical Methods}
\label{sec:numer}

We describe the governing equations and simulation set-up of our
two-component $\psiDM$ simulations.

\subsection{Governing equations}
\label{subsec:gov_eq}

In a two-component $\psiDM$ scenario, dark matter consists of one major
and one minor components distinguished by their contributions to the total
dark matter mass density. The particle masses associated with the major and minor
components are denoted by $\mM$ and $\mm$, respectively.
Each component can be described by a separate wave function coupled only by gravity.
The original wave function is real and satisfies the Klein-Gordon equation
under the influence of gravity \citep[see Appendix A of][]{Zhang2017PhRvDa}.
In the non-relativistic limit,
the wave function can be expressed as $\operatorname{Re}[\psi\exp({\rm i}\frac{mc^2}{\hbar} t)]$,
where $\psi$ is the non-relativistic wave function evolving on a slow time scale,
$c$ is the speed of light, and $\hbar$ is the reduced Planck constant.
The non-relativistic mass density is proportional
to $\bigl\langle\{\operatorname{Re}[\psi \exp({\rm i}\frac{mc^2}{\hbar} t)]\}^2\bigr\rangle$,
and the time average $\bigl\langle...\bigr\rangle$ removes the fast mass oscillation,
yielding the mass density as $|\psi|^2$.
For the two-component extension of the matter field in \cite{Zhang2017PhRvDa},
the mass density is proportional to
$\bigl\langle\{\operatorname{Re}[\psi_{\rm major}\exp({\rm i}\frac{\mM c^2}{\hbar} t)]\}^2+\{\operatorname{Re}[\psi_{\rm minor}\exp({\rm i}\frac{\mm c^2}{\hbar} t)]\}^2\bigr\rangle$,
where $\psi_{\rm major/minor}$ is the non-relativistic wave function for each component,
yielding the total mass density as $|\psi_{\rm major}|^2+|\psi_{\rm minor}|^2$.
Therefore, for the two-component $\psiDM$ in the non-relativistic limit,
the governing equations are given by
\be
{\rm i}\hbar\frac{\partial\psi_{\rm major}}{\partial t}=-\frac{\hbar^2}{2\mM}\nabla^2\psi_{\rm major}+\mM\Phi\psi_{\rm major},
\label{eq:schrodinger1}
\ee
\be
{\rm i}\hbar\frac{\partial\psi_{\rm minor}}{\partial t}=-\frac{\hbar^2}{2\mm}\nabla^2\psi_{\rm minor}+\mm\Phi\psi_{\rm minor},
\label{eq:schrodinger2}
\ee
\be
\nabla^2\Phi=4\pi G ( |\psi_{\rm major}|^2 + |\psi_{\rm minor}|^2 ),
\label{eq:poisson}
\ee
where $G$ is the gravitational constant, and
$\Phi$ is the common gravitational potential. Note that the mass density associated with
each component is given by $\rho_{\rm major/minor} = |\psi_{\rm major/minor}|^2$
and the total mass density is $\rho_{\rm total} = \rho_{\rm major} + \rho_{\rm minor}$.

\subsection{Simulation set-up}
\label{subsec:simu_setup}

\begin{table*}
\centering
\caption{Simulation set-up.}
\label{tab:SimulationSetup}
\begin{tabular}{lccc}
\hline
Model                                                                                 & \multicolumn{1}{c}{$\mratio=3:1$}                     & \multicolumn{1}{c}{$\mratio=1:3$}              & Single-component       \\ \hline
\begin{tabular}[c]{@{}l@{}}Major-component mass fraction\end{tabular}                 & \multicolumn{1}{c}{$75\%$}                            & \multicolumn{1}{c}{$75\%$}                     & $100\%$                \\
\begin{tabular}[c]{@{}l@{}}Minor-component mass fraction\end{tabular}                 & \multicolumn{1}{c}{$25\%$}                            & \multicolumn{1}{c}{$25\%$}                     & -                      \\
\begin{tabular}[c]{@{}l@{}}Major-component particle mass ($\mM$)\end{tabular}         & \multicolumn{1}{c}{$1 \times 10^{-22} \eV$}           & \multicolumn{1}{c}{$1 \times 10^{-22} \eV$}    & $1 \times 10^{-22} \eV$\\
\begin{tabular}[c]{@{}l@{}}Minor-component particle mass ($\mm$)\end{tabular}         & \multicolumn{1}{c}{$\frac{1}{3} \times 10^{-22} \eV$} & \multicolumn{1}{c}{$3 \times 10^{-22} \eV$}    & -                      \\
Starting redshift                                                                     & \multicolumn{1}{c}{$z=3200$}                          & \multicolumn{1}{c}{$z=3200$}                   & \multicolumn{1}{c}{$z=3200$}\\
Ending redshift                                                                       & \multicolumn{1}{c}{$z=0$}                             & \multicolumn{1}{c}{$z \sim 1 \textrm{--} 2$} & $z=0$                    \\
Initial condition                                                                     & \multicolumn{3}{c}{Same CDM power spectrum and spatial distribution}                                                            \\
Box size                                                                              & \multicolumn{1}{c}{$1.4 \Mpch$}                       & \multicolumn{1}{c}{$1.4 \Mpch$}                & \multicolumn{1}{c}{$1.4 \Mpch$}\\
Base-level resolution                                                                 & \multicolumn{1}{c}{$5.47 \kpch$}                      & \multicolumn{1}{c}{$1.37 \kpch$}               & $5.47 \kpch$           \\
Maximum resolution                                                                    & \multicolumn{1}{c}{$42.7 \pch$}                       & \multicolumn{1}{c}{$10.7 \pch$}                & $42.7 \pch$            \\ \hline
\end{tabular}
\end{table*}

We conduct cosmological simulations to investigate the non-linear
structure formation of the two-component $\psiDM$ model.
Throughout this work, we assume the major and minor components take
$75$ and $25$ per cent of the total dark matter mass, respectively.
We explore two different dark matter particle mass ratios,
$\mratio=3:1$ and $\mratio=1:3$,
where we fix $\mM=1\times10^{-22}\eV$ and choose $\mm$
to be either $\frac{1}{3}\times10^{-22}$ or $3\times10^{-22}\eV$.
For comparison, we also conduct single-component simulations with the same
particle mass as the major component (i.e. $10^{-22}\eV$).
Table~\ref{tab:SimulationSetup} lists the simulation set-up.

This work focuses on the properties of individual halo, which should be
insensitive to the initial power spectrum as long as it approaches the CDM
spectrum above the Jeans length and exhibits a strong cut-off below the
Jeans length.
Accordingly, we adopt the CDM power spectrum at an initial
redshift $z=3200$. The very high initial $z$ ensures that powers below the
Jeans length are significantly suppressed by quantum pressure at low $z$.
The results are insensitive to the exact value of initial $z$
and should be similar to simulations starting from typical $z\sim100$
with the two-component $\psiDM$ initial conditions
(though the tool for constructing such initial conditions is not currently available).
We also assume the initial density perturbations of the two components are perfectly
in phase. The three-dimensional realizations are generated by {\textsc{MUSIC}}
\citep{Hahn2011}. For $\mratio=3:1$, we conduct three runs
with different realizations to increase the sample size. We then apply
the same three initial conditions to both the $\mratio=1:3$ and
single-component runs.

We use $\gamer$ \citep{Schive2018}, a GPU-accelerated adaptive mesh refinement (AMR)
code, for our simulations. The original code only supports a single particle mass
\citep{Schive2014a} and we extend it to two components by solving
Eqs (\ref{eq:schrodinger1})--(\ref{eq:poisson}).
\sref{app:code_verify} provides the verification of the accuracy of our two-component implementation.
The periodic comoving box has a width of $1.4 \Mpch$, where $h=0.6732$ is the
Hubble parameter. The matter density parameter is $\Omega_m=0.3158$.
The simulations end at $z=0$ for both
$\mratio=3:1$ and the single-component runs but only reach $z=1.89$ for
$\mratio=1:3$ due to its much higher computational demands.

It is crucial to properly resolve the de Broglie wavelength in
$\psiDM$ simulations since we are evolving wave functions in
Eqs. (\ref{eq:schrodinger1}) and (\ref{eq:schrodinger2}).
It is especially challenging for $\mratio=1:3$ due to the short wavelength
associated with the minor component. To this end, we adopt both the
L\"{o}hner's error estimator \citep{Schive2014a} and mass density as the
AMR grid refinement criteria, the latter of which assures the central
solitons are always well resolved. The base-level resolution is $5.47\kpch$
for $\mratio=3:1$ and $1.37\kpch$ for $\mratio=1:3$, both with seven
refinement levels, achieving maximum resolution of $42.7$ and $10.7\pch$,
respectively. The data analysis and visualization are done with \texttt{yt}
\citep{yt}.

\section{Results}
\label{sec:resul}

We describe the large-scale structures of the entire simulation box first
followed by the structures of individual haloes and solitons.

\subsection{Large-scale structures}
\label{subsec:resultsoverall}

\begin{figure}
\centering
\includegraphics[width=\columnwidth]{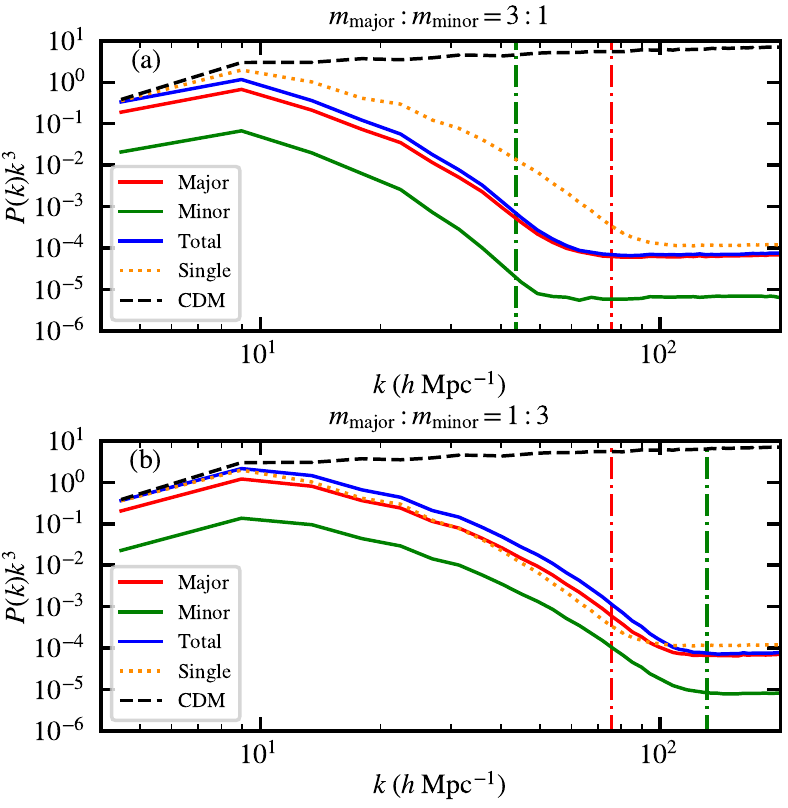}
\caption{
Density power spectra of the two-component model (red, green, and blue solid)
at $z=10$ compared to the single-component (orange dotted) and CDM (black dashed) models.
Panels (a) and (b) are for $\mM:\mm=3:1$ and $1:3$, respectively.
The CDM case shares the same initial power spectrum as $\psiDM$
but evolves as $P(k) \propto (1+z)^{-2}$ in the entire range of $k$ probed here.
The red and green dash-dotted vertical lines show the Jeans wavenumbers of the
major and minor components at $z=10$, respectively. The lowest-k mode of both the
two-component total density and single component follow the CDM evolution,
while larger-k modes are suppressed relative to CDM. In addition, the
total density power spectrum of $\mratio=3:1$ ($1:3$) is lower (higher)
than that of single component due to the stronger (weaker)
quantum pressure from the minor component.
}
\label{fig:PS}
\end{figure}

\begin{figure}
\centering
\includegraphics[width=\columnwidth]{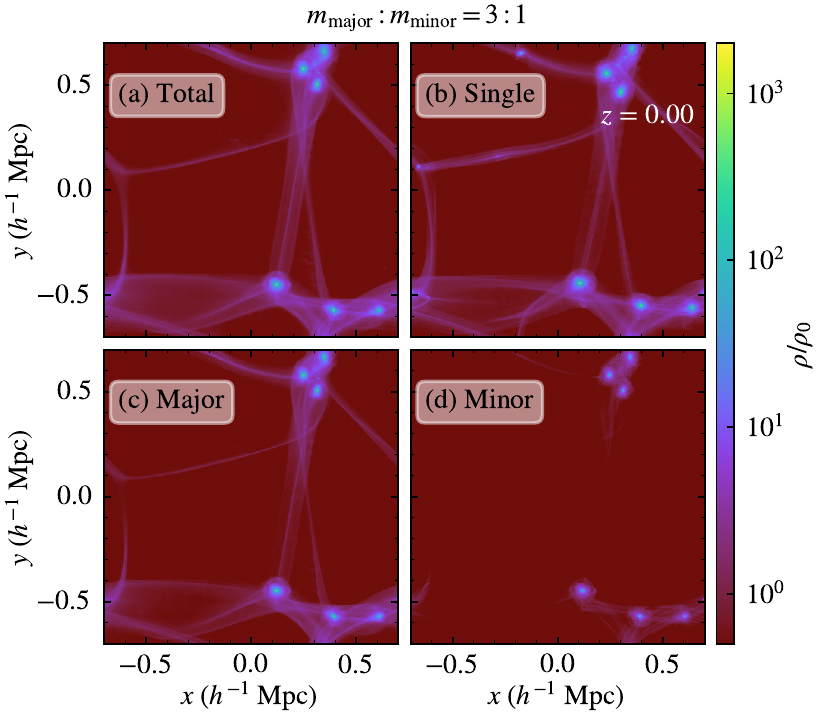}
\caption{
Projected density of the whole simulation box for $\mratio=3:1$ at $z=0$.
Panels (a), (c), (d) show the total, major-, and minor-component densities, respectively.
For comparison, we also plot the single-component result in panel (b).
The spatial distribution of both filaments and massive haloes are very
similar between the two- and single-component simulations, while there are
fewer low-mass haloes in the two-component case due to the stronger
quantum pressure resulting from the minor component.
Also note that the major-component haloes overlap well with their
minor-component counterparts, suggesting that every halo in our
two-component simulations is composed of both components.
}
\label{fig:3to1_GP}
\end{figure}

\begin{figure}
\centering
\includegraphics[width=\columnwidth]{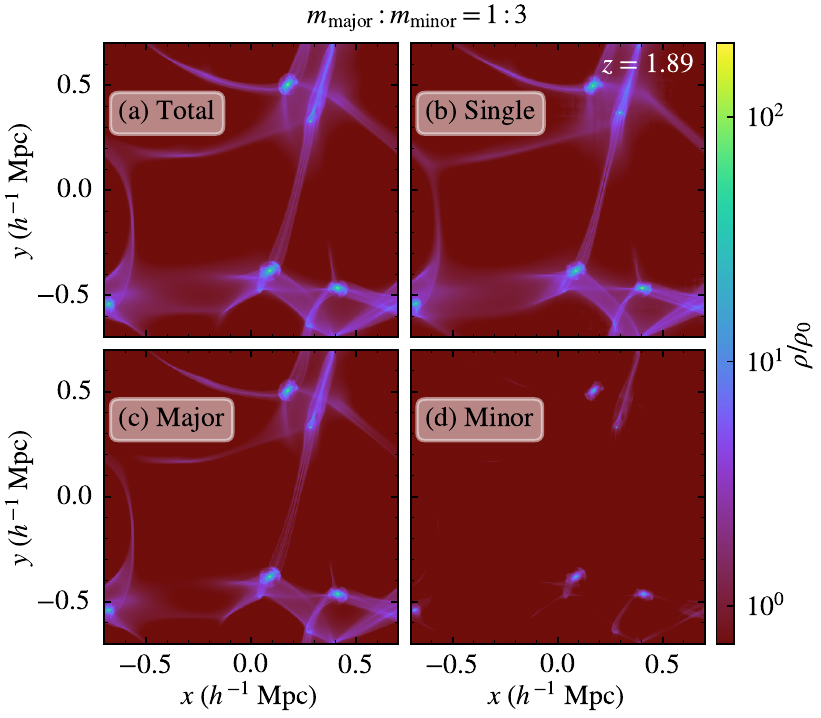}
\caption{
Same as \fref{fig:3to1_GP} except that here we show the results of $\mratio=1:3$
at $z=1.89$. Similar to $\mratio=3:1$, the spatial distribution of both filaments and haloes are very
similar between the two- and single-component simulations and the
major- and minor-component haloes overlap well with each other.
}
\label{fig:1to3_GP}
\end{figure}

\fref{fig:PS} compares the density power spectra of two-component,
single-component, and CDM models at $z=10$. The lowest-k mode of both the
two-component and single-component models follow the CDM evolution,
while larger-k modes are suppressed relative to CDM.
The level of suppression is $\mratio=3:1 > \mathrm{single~component} > \mratio=1:3$,
as expected since the Jeans wavenumber $k_{\rm J}=[6/(1+z)]^{1/4}(H_0 m/\hbar)^{1/2}$
becomes smaller for smaller $\mm$.

Figs. \ref{fig:3to1_GP} and \ref{fig:1to3_GP} show the projected density of the
whole simulation box for $\mratio=3:1$ and $1:3$, respectively,
normalized to the mean dark matter density of the universe $\rho_0$.
The single-component results are also shown for comparison.
The large-scale structures, including both filaments and massive haloes,
are very similar between the two- and single-component cases,
which is expected because they share the same initial condition
and the lowest-k mode is well above the Jeans scale. There are however
fewer low-mass haloes in $\mratio=3:1$ due to the stronger quantum pressure
resulting from a lighter minor component. Also note that for both
$\mratio=3:1$ and $1:3$, the major-component haloes overlap well with
their minor-component counterparts, suggesting that every halo in our
two-component simulations is composed of both components.

\subsection{Individual haloes and solitons}

We show the properties of individual haloes and solitons for the $\mratio=3:1$
simulations first followed by the $\mratio=1:3$ case.

\subsubsection{$\mratio=3:1$}
\label{subsec:results3to1}

\begin{figure}
\centering
\includegraphics[width=\columnwidth]{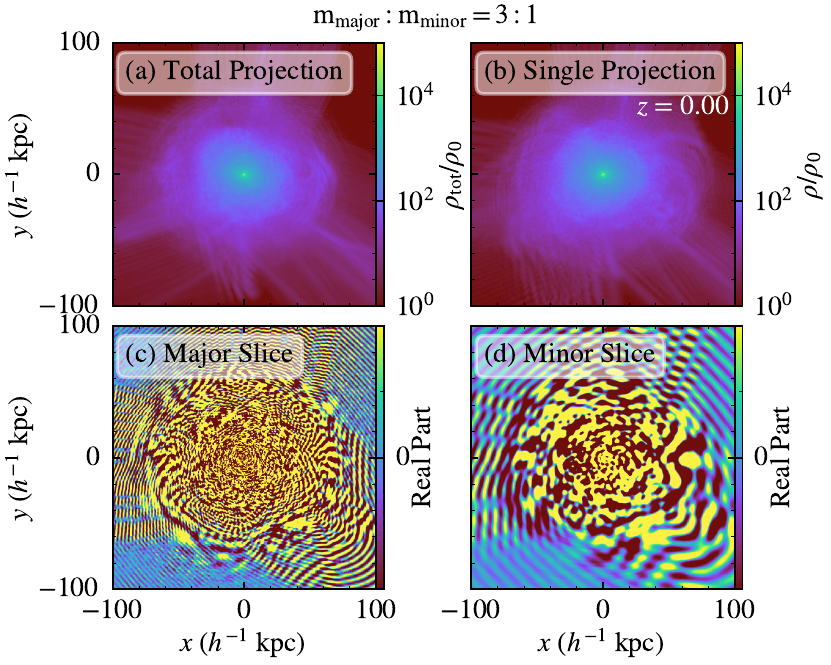}
\caption{
Overall structure of a representative halo in the $\mratio=3:1$ runs
at $z=0$. The halo mass is $\Mh \sim 2.1\times10^{10}\Msun$ and the
virial radius is $\sim 50\kpch$.
Panel (a) shows the projected total density of the two-component halo,
very similar to its single-component counterpart (b).
Panels (c) and (d) show the real parts of the wave functions of the
major and minor components, respectively, on a thin slice through the
densest cell. The wave functions are plane waves outside the virial radius
due to inflows and become isotropic inside the virialized halo.
The wavelength of the minor component is about three times longer than that
of the major component, in agreement with $\mratio=3:1$.
}
\label{fig:3to1_HP}
\end{figure}

\begin{figure}
\centering
\includegraphics[width=\columnwidth]{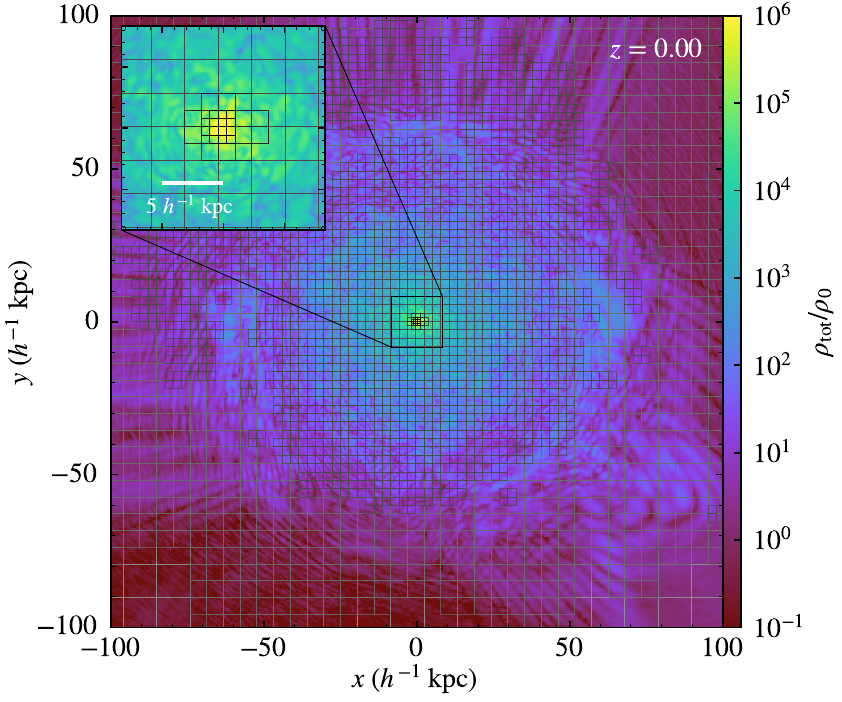}
\caption{
Example of the AMR grid structure. It shows the total density on a thin slice
through the centre of the same halo in \fref{fig:3to1_HP}, with AMR grids annotated.
Each grid represents $16\times16$ cells, with seven refinement levels,
achieving maximum resolution of $42.7\pch$. The inset shows the zoomed-in
view of the central high-density region of the halo. Thanks to AMR,
both the central soliton (see also Figs. \ref{fig:3to1_SP} and \ref{fig:3to1_P})
and the density granulation throughout the halo are well resolved.
}
\label{fig:3to1_HS}
\end{figure}

\begin{figure}
\centering
\includegraphics[width=\columnwidth]{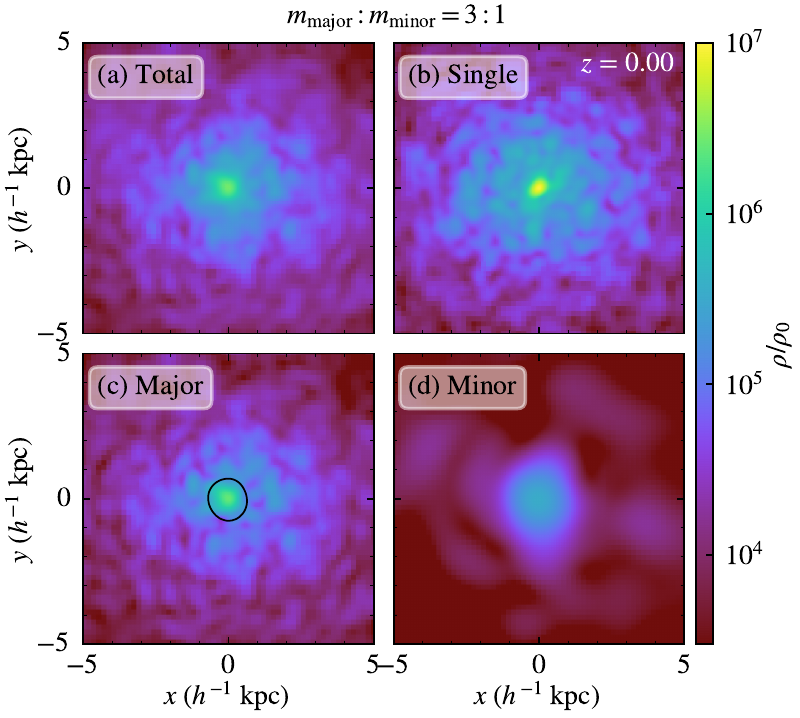}
\caption{
Zoomed-in view of the central high-density region of the same halo
in \fref{fig:3to1_HP}, showing the projected density in a slab of dimensions
$10\kpch \times 10\kpch \times 3.9\kpch$ centred at the soliton.
We plot the total density of the two-component halo (a),
the associated major-component (c) and minor-component (d) densities
(so $\textrm{(a)}=\textrm{(c)}+\textrm{(d)}$), and the single-component halo (b).
Solitons and density granules are manifest in both the two- and
single-component haloes.
In comparison to the single-component halo, in the two-component case
the soliton peak density is significantly lower,
the transition from soliton to halo is smoother (see also \fref{fig:3to1_P}),
and the density granules are slightly blurrier due to the presence of a
longer-wavelength minor component.
Solitons appear in both the major and minor components.
The soliton radius and granule size of the minor component are about three
times larger than that of the major component, in line with $\mratio=3:1$.
The solid line in panel (c), defined as the isodensity contour where the
projected minor-component density drops by half, indicates the boundary of the
minor-component soliton. It shows that the major- and minor-component
solitons are roughly concentric (see also \fref{fig:3to1_RD}).
}
\label{fig:3to1_SP}
\end{figure}

\begin{figure}
\centering
\includegraphics[width=\columnwidth]{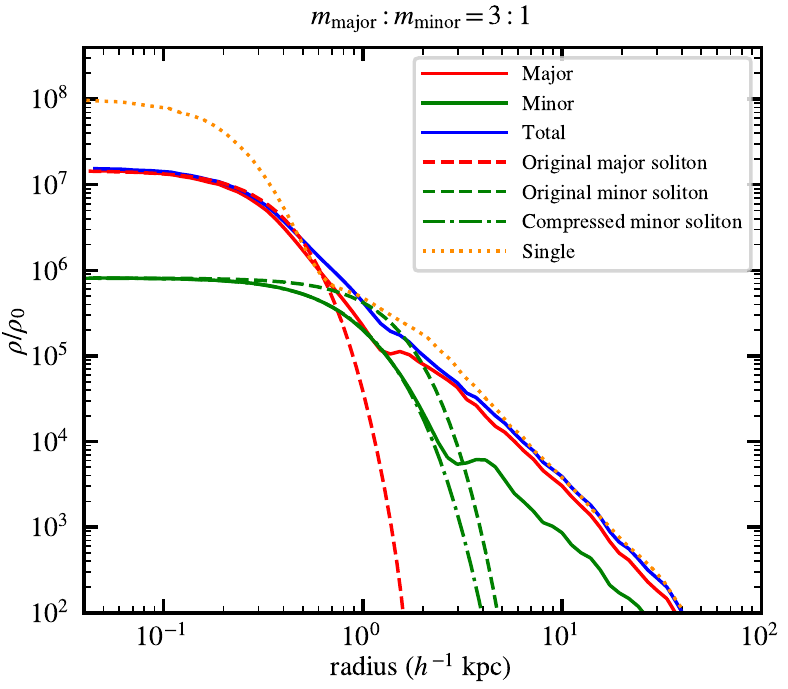}
\caption{
Density profiles of the same halo in \fref{fig:3to1_HP}, centred at the
location of the maximum density for each case.
The solid and dotted lines represent respectively the two- and single-component
haloes, the total density profiles of which at large radii are very similar.
However, the peak density of the two-component soliton is about five times lower.
The dashed lines are the analytical soliton profiles,
\eref{eq:soliton}, without considering any external potential,
while the dash-dotted line is the \emph{compressed} soliton profile for the
minor component considering the major-component soliton (red dashed line)
as an external potential.
As can be seen, the major-component cored profile can be well fitted by the
original soliton solution, while the minor-component cored profile is
better described by the compressed soliton solution.
The less prominent major-component soliton together with the presence of an
extended minor-component soliton lead to a much smoother soliton-to-halo
transition (at $\sim 1\kpch$) in the two-component halo with $\mratio=3:1$
(see also \fref{fig:RC}).
}
\label{fig:3to1_P}
\end{figure}

\begin{figure}
\centering
\includegraphics[width=\columnwidth]{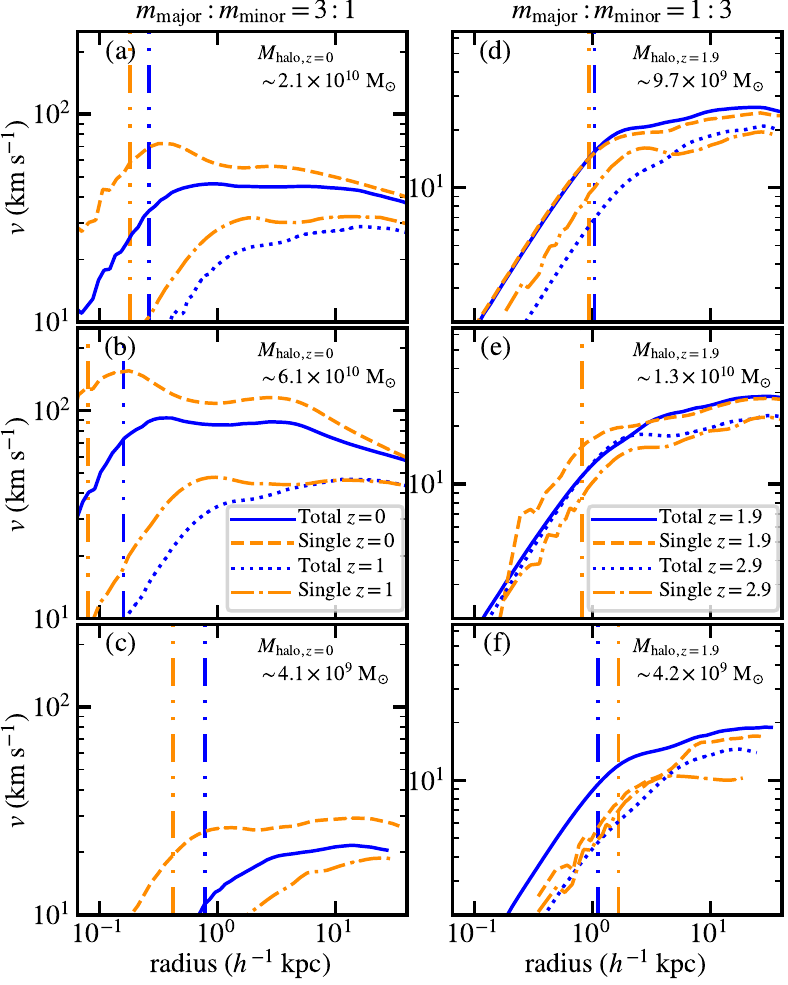}
\caption{
Rotation curves of the two-component haloes.
Panels (a)--(c) show different haloes for $\mratio=3:1$
and panels (d)--(f) show different haloes for $\mratio=1:3$.
Panels (a) and (d) are the same haloes in Figs. \ref{fig:3to1_HP} and \ref{fig:1to3_HP}, respectively.
The solid blue lines represent the two-component haloes at their final redshifts
(i.e. $z=0$ for $\mratio=3:1$ and $z=1.89$ for $\mratio=1:3$).
The dashed orange lines show the single-component counterparts.
The vertical dash-double-dotted lines mark the core radii of both the two- and single-component
haloes at their final redshifts.
The annotated $\Mh$ at corners of panels indicate their two-component halo masses
at their final redshifts.
The dotted blue and dash-dotted orange lines show the two- and single-component haloes at
earlier redshifts ($z=1$ for $\mratio=3:1$ and $z=2.94$ for $\mratio=1:3$).
The `bump' in the rotation velocity, an imprint of the soliton, is less
prominent in $\mratio=3:1$ due to the smoother soliton-to-halo transition
(see also \fref{fig:3to1_P}) in comparison to the single-component counterpart.
In comparison, for $\mratio=1:3$, the rotation curves of
the two- and single-component haloes are similar due to the absence of a minor-component
soliton (see also \fref{fig:1to3_P}).
Note that the bumps in the rotation curves of panels (d)--(f) are less prominent and sometimes disappear
since solitons are less stable at higher redshifts.
The two-component halo at $z=1$ in panel (c) is not shown because its density is too low.
}
\label{fig:RC}
\end{figure}

\begin{figure}
\centering
\includegraphics[width=\columnwidth]{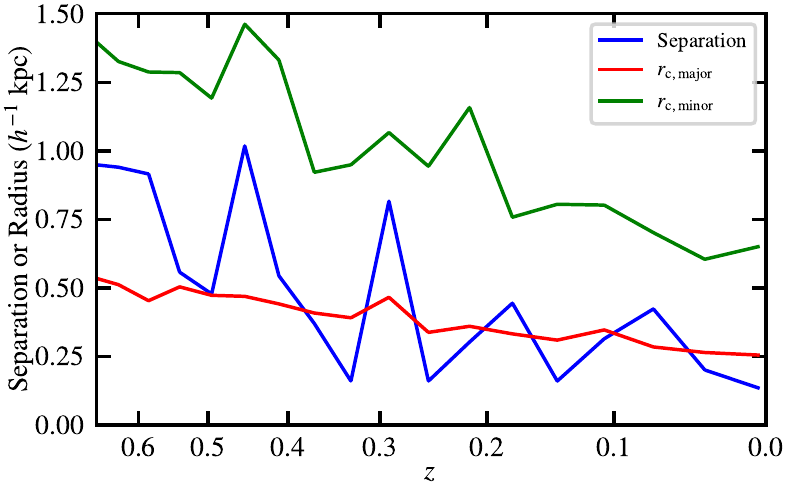}
\caption{
Separation distance between the major- and minor-component solitons at
different redshifts for $\mratio=3:1$ (blue).
For comparison, the red and green lines show the soliton core radii,
$\rcM$ and $\rcm$, satisfying $\rcm/\rcM \sim \mM/\mm = 3$.
The separation distance is always smaller than $\rcm$, suggesting that the
two solitons move together and are roughly concentric.
This separation is likely caused by the random walk of the major-component
soliton \citep{Schive2020}.
}
\label{fig:3to1_RD}
\end{figure}

\begin{figure}
\centering
\includegraphics[width=\columnwidth]{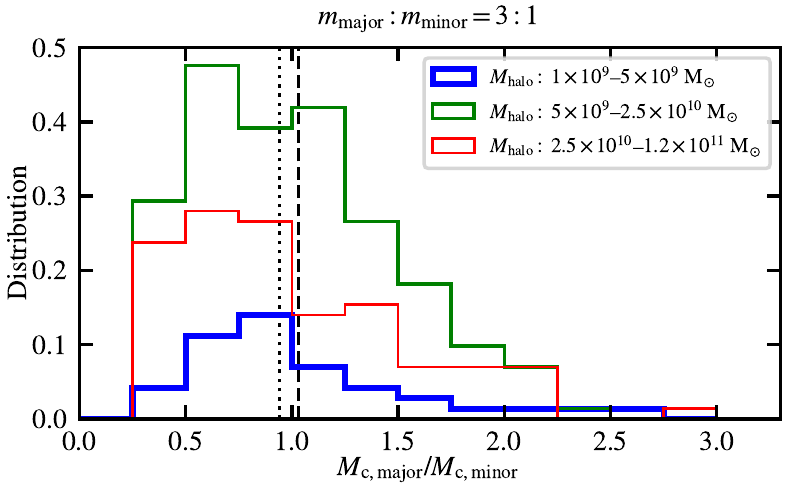}
\caption{
Distribution of the mass ratio between the major- and minor-component solitons
in the same haloes, $\McM/\Mcm$, for $\mratio=3:1$.
We include all haloes after $z<0.63$ in the
three realizations and divide them into three groups by the halo mass $\Mh$
(blue, green, and red). The distribution is normalized such that the
sum of the three areas equals unity.
The dashed and the dotted vertical lines show the average and
the median values of the whole distribution, respectively.
The individual average and median values of each group are all
within the $\pm4$ per cent range of the vertical dashed and dotted lines, respectively.
$\McM$ and $\Mcm$ are found to be comparable, with ratios
ranging mostly between $0.5 \textrm{--} 2.0$ and independent of $\Mh$.
}
\label{fig:3to1_SMRD}
\end{figure}

Here the particle masses of the major and minor components are
$\mM=1\times10^{-22}\eV$ and $\mm=\frac{1}{3}\times10^{-22}\eV$, respectively.
\fref{fig:3to1_HP} shows the density distribution of a representative halo
in our simulations, with a halo mass $\Mh \sim 2.1\times10^{10}\Msun$ and
a virial radius $\sim 50\kpch$.
The overall structure of the two-component halo closely mimics its
single-component counterpart. The wave functions outside the virial radius
exhibit plane-wave features due to inflows; by contrast, the wave functions
become isotropic inside the halo as a result of virialization. The wavelength
is slightly shorter toward the halo centre due to the increment of velocity.
In addition, the wavelength of the minor component is about three times
longer than that of the major component, in agreement with $\mratio=3:1$.

\fref{fig:3to1_HS} shows the same halo in \fref{fig:3to1_HP} but with AMR grids
annotated to highlight the high resolution achieved. The mean spatial
resolution within the halo is $170\pch$ and the peak resolution is
$42.7\pch$ in the central high-density region. In comparison, the
characteristic radius of the central soliton (see \fref{fig:3to1_P}) is
about several hundreds of $\pch$. Clearly, both the soliton and the
density granulation throughout the halo are well resolved.

Single-component $\psiDM$ haloes feature a central, dense soliton surrounded
by density granules on the de Broglie scale. \fref{fig:3to1_SP} shows a zoomed-in
view of the central high-density region of the same halo in \fref{fig:3to1_HP},
demonstrating that both the soliton and density granules are still manifest
in a two-component halo. However, we find several key differences between
the single- and two-component cases with $\mratio=3:1$ due to the presence
of a longer-wavelength minor component.
The density of the two-component soliton is less dense, resulting in a
smoother transition from soliton to its surrounding halo. In addition,
the density granules in a two-component halo are slightly blurrier.
\fref{fig:3to1_SP} also depicts separately the density distribution of the
major and minor components, showing that solitons appear in both components.
The characteristic length scales of the minor-component soliton and density
granules are about three times larger than that of the major component,
consistent with $\mratio=3:1$.
Furthermore, the major- and minor-component solitons are roughly concentric,
suggesting that the soliton total density can be seen as the superposition
of the individual soliton in each component.
In what follows, we analyse these interesting findings in more detail.

\fref{fig:3to1_P} shows the radial density profiles of the same halo in \fref{fig:3to1_HP}.
It confirms that the peak density of the two-component soliton is about
a factor of five lower, making the solitonic structure less prominent in comparison
to the single-component counterpart.
On the other hand, the total density profiles at large radii are very consistent
between the two cases.
The major-component cored profile can be well fitted by the soliton
solution \citep{Schive2014a},
\be
\rho(r)=\frac{1.9(m/10^{-23}\eV)^{-2}(\mathit{r}_{\rm c}/\kpc)^{-4}}{[1+9.1\times10^{-2}(r/r_{\rm c})^2]^8}\Msun \pc^{-3},
\label{eq:soliton}
\ee
where $r_{\rm c}$ is the core radius where density drops to one-half of its peak value\footnote{
Throughout this paper, we use the term ``core radius $r_{\rm c}$'' to refer to
the radius where the shell-average density in the simulation data is one-half of the peak density,
but not the core radius derived from the fitted original or compressed soliton solution.}.
On the contrary, the minor-component cored profile cannot be well fitted by
\eref{eq:soliton} since this component is more extended (due to a
three-times lighter particle mass) and thus subject to the gravitational
attraction of the interior major-component soliton.
As a result, we find that the minor-component cored profile can be well fitted
by a \emph{compressed} soliton solution taking into account the
major-component soliton as an external potential.
The core radii of the major, minor, and single components are
$0.25, 0.65$ and $0.18\kpch$, respectively.
As has been noted, the total density profile of the two-component soliton
can be seen as the superposition of the individual soliton in each component
with different length scales and shapes.
In consequence, the less prominent major-component soliton together with the
presence of an extended minor-component soliton lead to a much smoother
soliton-to-halo transition in a two-component halo with $\mratio=3:1$.

Importantly, this smoother soliton-to-halo transition in the density profile
will affect the associated rotation curves, as shown in \fref{fig:RC}.
The `bump' in the rotation velocity, an imprint of the massive and
compact soliton, becomes much less prominent in the $\mratio=3:1$ halo
as compared to its single-component counterpart.
This new finding may have a significant impact since the predicted bump
in the rotation velocity of the single-component $\psiDM$ scenario is
in tension with observations and posing a serious challenge against
$\psiDM$ \citep[e.g.][]{Bar2022}.

\fref{fig:3to1_SP} suggests that the major- and minor-component solitons
in a $\mratio=3:1$ halo can coexist and are roughly concentric.
To investigate whether it is a general feature or just a coincidence,
in \fref{fig:3to1_RD} we plot the distance between the two solitons at
different redshifts. It demonstrates that the concentricity of the two solitons
is a general phenomenon for $\mratio=3:1$, for which the ratio of their core
radii is $\rcm/\rcM \sim \mM/\mm = 3$ and their maximum separation is
always smaller than $\rcm$. This separation is likely caused by the random
walk of the major-component soliton \citep{Schive2020}.

Although for $\mratio=3:1$ the minor-component solitons have lower peak density
in general (\fref{fig:3to1_P}), they also have larger radii (\fref{fig:3to1_RD}).
So it remains unclear whether the major- or minor-component soliton is more
massive for a given halo mass, and how does that depend on the halo mass.
To address this question, we show in \fref{fig:3to1_SMRD} the distribution of
the mass ratio between the major- and minor-component solitons in same haloes,
$\McM/\Mcm$, where the soliton mass
$M_{\rm c}$ is defined as the enclosed mass within the core radius $r_{\rm c}$.
It reveals that the masses of the two solitons are comparable, with
ratios ranging mostly between $0.5 \textrm{--} 2.0$ and independent of the halo mass.
Note that this is a new finding different from
the soliton thermodynamic expectation.
Theoretically, the soliton mass is proportional to $m^{-1}$
for a given halo mass \citep{Schive2014PhRvL}.
With the same virial temperature,
the mass ratio between the major- and minor-component solitons
is expected to be $1:3$ in this case.
However, the ratios we found in the simulations are $\sim 1:1$.

\subsubsection{$\mratio=1:3$}
\label{subsec:results1to3}

\begin{figure}
\centering
\includegraphics[width=\columnwidth]{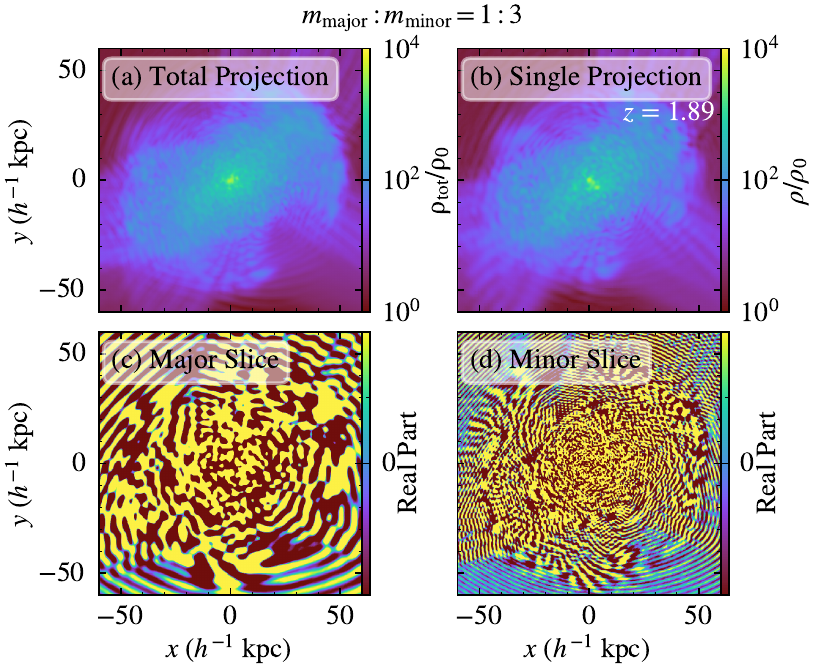}
\caption{
Overall structure of a representative halo in the $\mratio=1:3$ run
at $z=1.89$. The halo mass is $\Mh \sim 9.7\times10^{9}\Msun$, corresponding
to a comoving virial radius of $\sim 47\kpch$.
The general features of this halo are similar to the $\mratio=3:1$ case
shown in \fref{fig:3to1_HP}.
Panel (a) shows the projected total density of the two-component halo,
very similar to its single-component counterpart (b).
Panels (c) and (d) show the real parts of the wave functions of the
major and minor components, respectively, on a thin slice through the
densest cell. The wave functions are plane waves outside the virial radius
due to inflows and become isotropic inside the virialized halo.
The wavelength of the minor component is about three times shorter than that
of the major component, in agreement with $\mratio=1:3$.
}
\label{fig:1to3_HP}
\end{figure}

\begin{figure}
\centering
\includegraphics[width=\columnwidth]{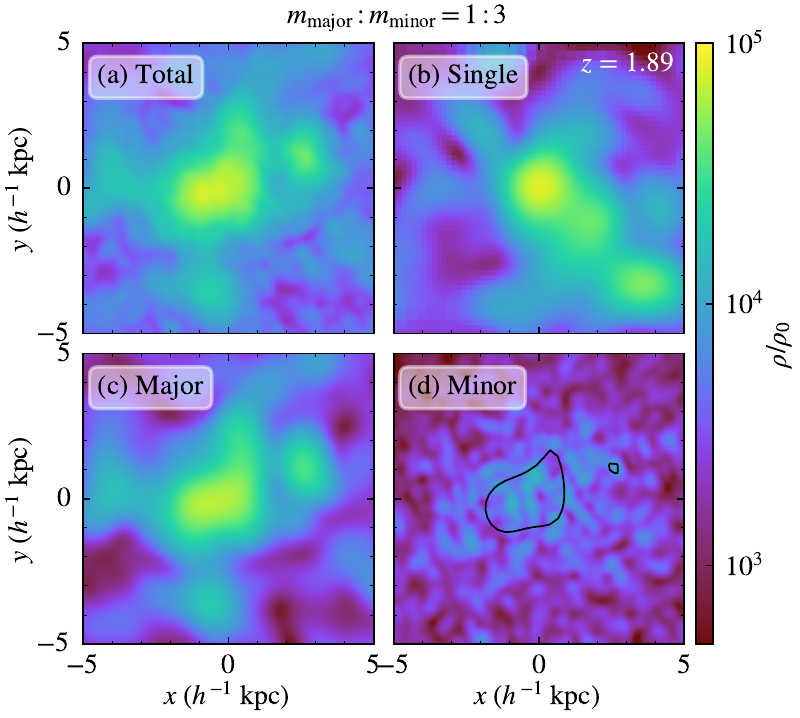}
\caption{
Zoomed-in view of the central high-density region of the same halo
in \fref{fig:1to3_HP}, showing the projected density in a slab of dimensions
$10\kpch \times 10\kpch \times 5.98\kpch$ centred at the central soliton.
We plot the total density of the two-component halo (a),
the associated major-component (c) and minor-component (d) densities
(so $\textrm{(a)}=\textrm{(c)}+\textrm{(d)}$), and the single-component halo (b).
A prominent soliton can be clearly seen in the \emph{total} density of both
the two- and single-component haloes.
However, for the former, the soliton only forms in the major
component but not in the minor component.
To highlight this finding, we plot in panel (d) the boundary of the major-component soliton
(solid line), defined as the isodensity contour where the projected major-component density drops by half.
Clearly, there is no minor-component high-density clump inside the
major-component soliton.
This result is distinctly different from the $\mratio=3:1$ case
(e.g. see \fref{fig:3to1_SP}), where the major- and minor-component solitons
can coexist and are roughly concentric.
Density granulation appears in both components,
the size of which is about three smaller in the minor component,
consistent with $\mratio=1:3$.
As a result, compared to the single-component case, the total density
granulation of the two-component halo exhibits more fine structures
due to the presence of a shorter-wavelength minor component.
}
\label{fig:1to3_SP}
\end{figure}

\begin{figure}
\centering
\includegraphics[width=\columnwidth]{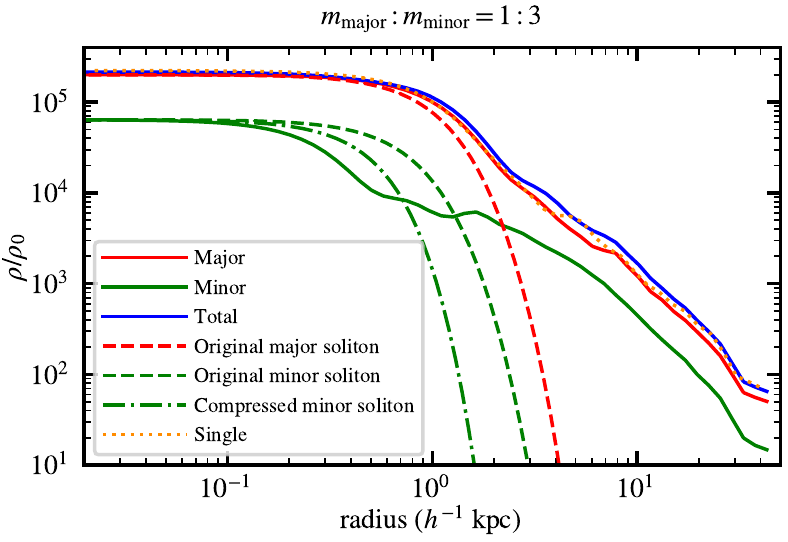}
\caption{
Density profiles of the same halo in \fref{fig:1to3_HP}, similar to \fref{fig:3to1_P}
except that here is for $\mratio=1:3$. The total density profiles of the
two- and single-component haloes roughly follow each other not only at large radii
but also in their central cores, which can be well fitted by the analytical
soliton solution \eref{eq:soliton}.
The two-component
halo virial mass is $\sim 15$ per cent larger than the single-component halo,
likely attributed to the weaker quantum pressure associated with the minor component
with a higher particle mass.
In contrast to the $\mratio=3:1$ case, here the single- and two-component cases
have a similar soliton peak density on average and both exhibit a distinct
soliton-to-halo transition.
More importantly, although the minor-component profile still has a density
clump at the centre, it is \emph{not} a soliton because it is much lighter
(also referred to \fref{fig:1to3_SP}) and cannot be well fitted by the
soliton solution. See text for details.
}
\label{fig:1to3_P}
\end{figure}

\begin{figure}
\centering
\includegraphics[width=\columnwidth]{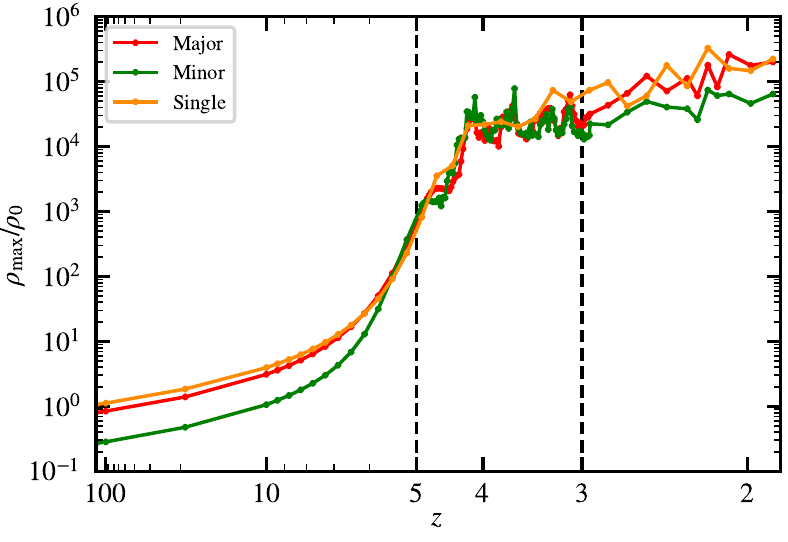}
\caption{
Evolution of the peak density of the major ($\dmaxM$; red) and minor
($\dmaxm$; green) components of the same $\mratio=1:3$ halo in \fref{fig:1to3_HP}.
We also plot the single-component counterpart (orange) for comparison.
For $z \gtrsim 10$, we have $\dmaxM/\dmaxm \sim 3$, in accordance with
the initial condition.
Subsequently, the minor component grows faster due to the weaker
quantum pressure and collapses at $z \sim 5$.
The major component also collapses shortly afterwards.
Both components then go through a violent relaxation process over
$5 \gtrsim z \gtrsim 3$, during which both the major- and minor-component
solitons form and disrupt continuously and stochastically.
After $z \lesssim 3$, the major-component soliton becomes more stable
and $\dmaxM$ starts to grow on average again;
meanwhile the minor-component soliton does not re-emerge and
$\dmaxm < \dmaxM$ thereafter.
The two vertical lines mark $z=5$ and $3$ that roughly separate the
three evolution stages mentioned above.
}
\label{fig:1to3_PDE}
\end{figure}

\begin{figure}
\centering
\includegraphics[width=\columnwidth]{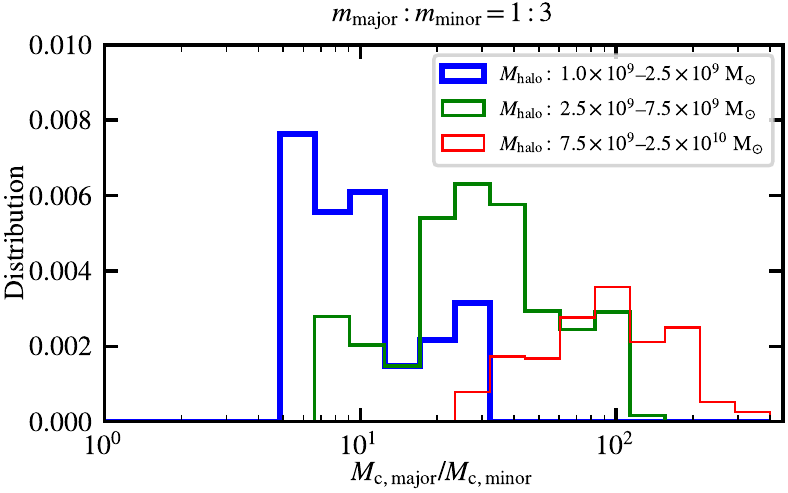}
\caption{
Distribution of the mass ratio between the major- and minor-component solitons
in the same haloes, $\McM/\Mcm$, for $\mratio=1:3$.
We include all haloes after $z<3$ in the
three realizations and divide them into
three groups by the halo mass $\Mh$ (blue, green, and red).
For the comparison purpose, we define the \emph{soliton} mass as the
enclosed mass within the half-density radius (i.e. $r_{\rm c}$) even when
there is no real minor-component soliton (see Figs. \ref{fig:1to3_SP}--\ref{fig:1to3_PDE}
and text for the related discussion).
The horizontal bins here are in log scale and the distribution is
normalized such that the sum of the three areas equals unity.
It shows that the mass ratio is always larger than $5$ and mostly
exceeds $\sim 10$, especially for massive haloes,
suggesting that the minor-component solitons have negligible effect
in most cases.
}
\label{fig:1to3_SMRD}
\end{figure}

In this case, the particle mass of the minor component is
$\mm=3\times10^{-22}\eV$, three times larger than that of the
major component, $\mM=1\times10^{-22}\eV$.
The time and length scales associated with the minor component are thus
three times smaller. As a result, the simulation can only reach $z=1.89$
because of the much higher spatial and temporal resolutions required to
properly resolve the minor component.
Nevertheless, we can already identify distinct features compared to
the $\mratio=3:1$ case.

\fref{fig:1to3_HP} shows the density distribution of a representative halo
with $\Mh \sim 9.7\times10^{9}\Msun$, corresponding to a comoving
virial radius of $\sim 47\kpch$.
In analogy to the $\mratio=3:1$ case shown in \fref{fig:3to1_HP}, the overall
structures of the two- and single-component haloes are very similar.
The wave functions exhibit plane-wave features outside the virial radius
but become isotropic inside the virialized halo. The wavelength
is slightly shorter toward the halo centre due to the increment of velocity.
The wavelength of the minor component is about three times
shorter than that of the major component, as expected from $\mratio=1:3$.

\fref{fig:1to3_SP} shows a zoomed-in view of the central high-density region of
\fref{fig:1to3_HP}. Both the soliton and density granulation are manifest in the
\emph{total} density of the two-component halo, similar to its
single-component counterpart. However, surprisingly, we find that for the
two-component case with $\mratio=1:3$ the soliton only forms in the
major component but not in the minor component (see panel (d)).
This result is distinctly different from the $\mratio=3:1$ case
(e.g. see \fref{fig:3to1_SP}), where the major- and minor-component solitons
can coexist and are roughly concentric.
By contrast, density granulation still appears in both the major and minor
components. As the images in panels (c) and (d) show,
the characteristic length scale of the minor-component
density granules is about three times smaller than that of the
major component, consistent with the theoretical expectation of wavelength scale
with the same velocity dispersion and $\mratio=1:3$.
As a result, comparing panels (a) and (b) shows that the total density
granulation of the two-component halo exhibits more fine structures
due to the presence of a shorter-wavelength minor component.

The discovery of the absence of a minor-component soliton with $\mratio=1:3$
is important and contradictory to some previous studies
\citep[e.g.][]{Luu2020}. In the following we provide a more detailed analysis
to strengthen this finding.
\fref{fig:1to3_P} plots the radial density profiles of the same halo in \fref{fig:1to3_HP}.
It shows that the total density profiles of the two- and single-component haloes
are similar at large radii, in analogy to the $\mratio=3:1$ case
(see \fref{fig:3to1_P}). However, unlike $\mratio=3:1$, here
the two- and single-component cases also have a similar soliton profile on average
and both exhibit a distinct soliton-to-halo transition.
More importantly, it confirms that although the minor-component profile
still exhibits a central density clump, it is \emph{not} a soliton.
Instead, this density clump is nothing but a slightly more massive granule,
as suggested by the facts that (i) its density contrast is much smaller
compared to a typical soliton, (ii) its peak density is only
about a factor of five higher than the peak density of surrounding granules, and (iii) it
is inconsistent with the soliton solution even after considering the
additional compression due to the gravity of the major-component soliton
(dash-dotted line). The absence of any soliton feature in panel (d) of
\fref{fig:1to3_SP} also reinforces this claim.
As a result, the total density profile is dominated by the major component
over the entire radius range.
The corresponding rotation curves are shown in \fref{fig:RC}.
The half-density radii of the major, minor, and single components are
$1.00, 0.27$ and $0.93\kpch$, respectively.

Figs. \ref{fig:1to3_HP}--\ref{fig:1to3_P} focus on the results at the end of a
$\mratio=1:3$ simulation, at which the minor-component soliton has
disappeared. But what about the earlier stage? Did the minor-component
soliton never form, or was it somehow disrupted at a later stage?
To address this question, we plot in \fref{fig:1to3_PDE} the evolution of the
peak density of the major and minor components, $\dmaxM$ and $\dmaxm$,
of the same $\mratio=1:3$ halo in \fref{fig:1to3_HP}.
It shows that for $z \gtrsim 10$, we have $\dmaxM/\dmaxm \sim 3$,
consistent with the initial condition (see \sref{subsec:simu_setup}).
Afterwards, the minor component starts to grow faster than the major
component due to the weaker quantum pressure associated with the larger
particle mass. The minor component collapses first at $z \sim 5$,
shortly after which the major component also collapses.
During $5 \gtrsim z \gtrsim 3$, both the major- and minor-component
solitons can form but are rather unstable and get destroyed constantly and
stochastically, suggesting that they are undergoing a violent
relaxation process.
In addition, $\dmaxM$ and $\dmaxm$ are comparable and both exhibit
large-amplitude oscillations during this period.
After $z \lesssim 3$, the major-component soliton becomes more stable and
$\dmaxM$ starts to grow on average again.
It is at this stage that the minor-component soliton cannot re-emerge.
This indicates that it is difficult for the minor-component soliton to form
in a \emph{hot} environment associated with the deep gravitational potential
of the major-component soliton.
We will discuss it in more detail in \sref{sec:discu}.

\fref{fig:1to3_SMRD} shows the distribution of the mass ratio between the major-
and minor-component solitons in the same haloes.
For the comparison purpose, we still define the soliton mass $M_{\rm c}$ as the
enclosed mass within the half-density radius $r_{\rm c}$ even when there is
no real minor-component soliton.
It shows that the mass ratio is always larger than $5$ and mostly
exceeds $\sim 10$, especially for massive haloes.
This suggests that the minor-component solitons (or massive granules when
there is no real soliton) in the $\mratio=1:3$ case have negligible effect
in most cases, distinctly different from the $\mratio=3:1$ scenario
(see \fref{fig:3to1_SMRD}). Moreover, for the minor component, the gravitational
potential is mostly external and determined by the major component,
and we find that the minor component is difficult to form a soliton in such an environment.
See \sref{sec:discu} for more discussion on this point.

\section{Discussion}
\label{sec:discu}

\begin{figure}
\centering
\includegraphics[width=\columnwidth]{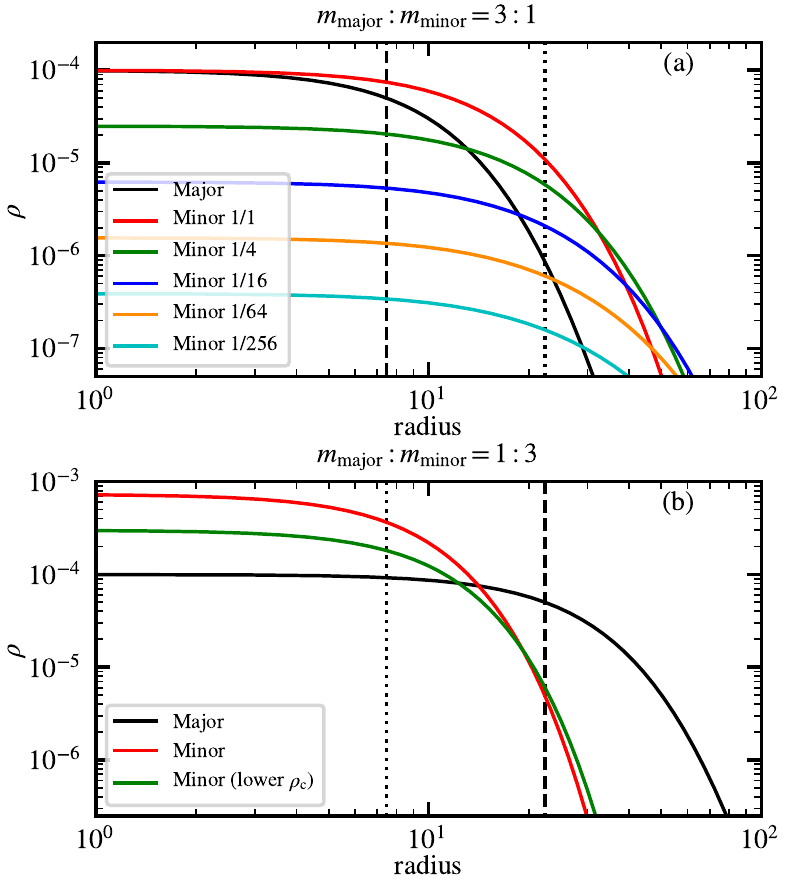}
\caption{
Analytical two-component soliton solutions.
The black lines are the major-component solitons given by the original
soliton solution \eref{eq:soliton}. The corresponding gravitational potential
is then considered as a static external potential for computing the
minor-component solitons.
The vertical dashed lines show the major-component
soliton radius $\rcM$.
The vertical dotted lines show
$3\rcM$ and $\rcM/3$ for $\mratio=3:1$ and
$\mratio=1:3$, respectively.
Panel (a) shows the $\mratio=3:1$ case, where the red, green, blue, orange,
and cyan lines show respectively the minor-component solitons with different
peak density ratios,
$\rho_{\rm c, major}$/$\rho_{\rm c, minor} = 1, 4, 16, 64, 256$.
With a lower $\rho_{\rm c, minor}$, the minor-component soliton radius increases
and approaches a constant ratio $\rcM/\rcm \sim 0.4$,
broadly consistent with the expected value for $\mratio=3:1$
(see also \fref{fig:3to1_RD}).
The mass ratio $\McM/\Mcm$, on the other hand,
can be greater or less than unity across a wide range of peak densities,
in agreement with our cosmological simulations (\fref{fig:3to1_SMRD}).
Panel (b) shows the $\mratio=1:3$ case, where we fix
$\rcM/\rcm = 3$ for the red line, leading to
$\rho_{\rm c, major}$/$\rho_{\rm c, minor} = 1/7.3$ and
$\McM/\Mcm = 3.7$.
The green line shows the minor-component soliton with a lower peak density,
where $\rho_{\rm c, major}$/$\rho_{\rm c, minor} = 1/3$ and
$\McM/\Mcm = 5.5$.
In comparison, the minor-component solitons in our cosmological simulations
either have a relatively lower density or cannot sustain at all.
See text for further discussion.
}
\label{fig:ATCSP}
\end{figure}

\begin{figure*}
\centering
\includegraphics[width=\textwidth]{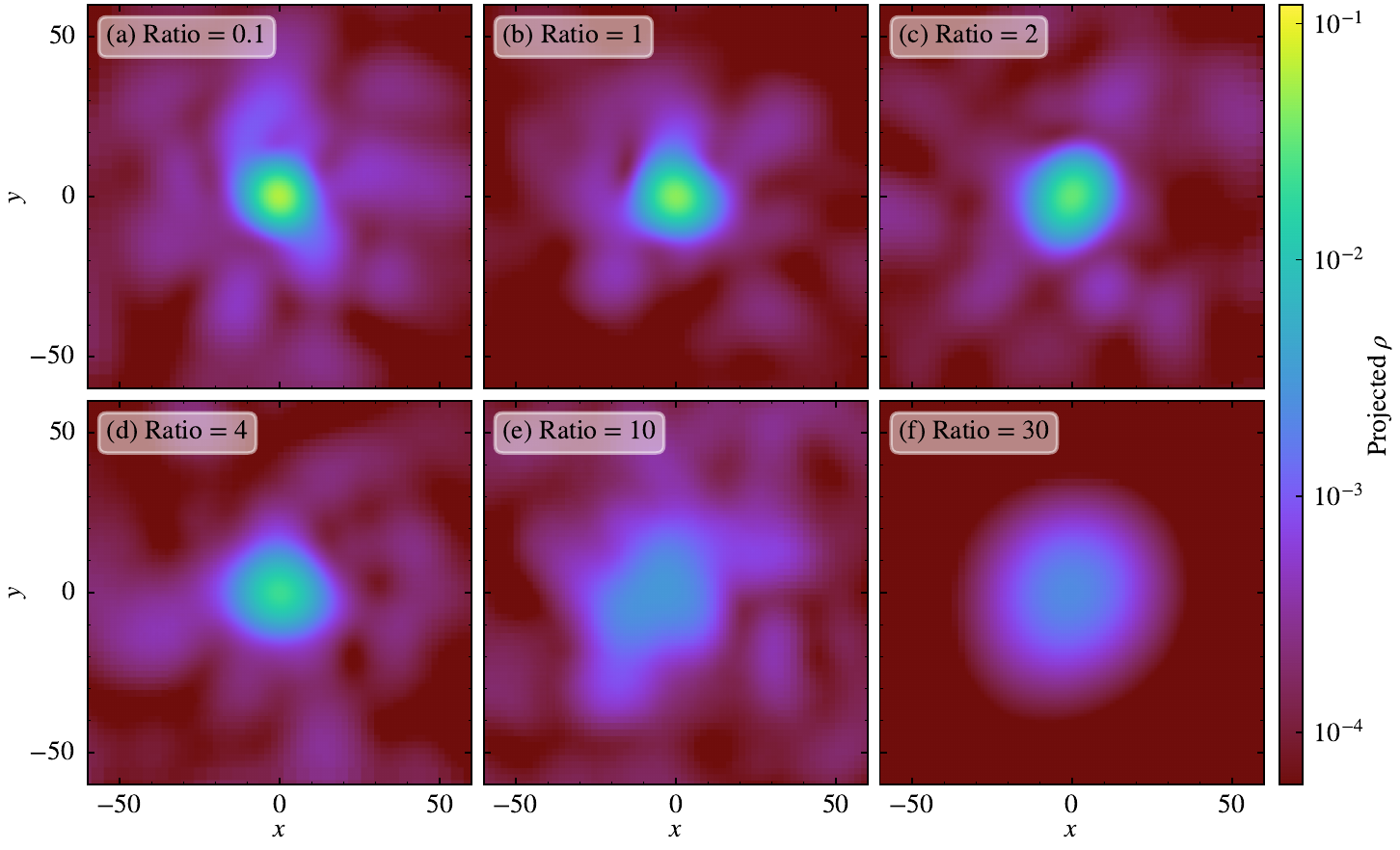}
\caption{
Projected density in the toy model described in \sref{subsec:toy_model},
with ten minor-component solitons merging in a static simple harmonic potential
mimicking a major-component soliton. This figure shows
the final relaxed state for investigating whether a single soliton can re-emerge
after merger. Different panels show the simulations with different energy ratios
between the simple harmonic potential energy and the self-gravitational
potential energy of all solitons in the initial conditions.
As the energy ratio increases (i.e. a deeper simple harmonic potential),
the central object in the final state becomes more diffusive and deviates
from the soliton solution (see \fref{fig:EPMPf}), especially when the
energy ratio exceeds $\sim 10$ (panels (e) and (f)).
Not only the soliton cannot form at the end, but it never forms
during the entire merging and relaxation process when the energy ratio is large.
It indicates that a soliton is difficult to form in a hot environment under
a deep potential. This toy model provides a plausible explanation for
why a minor-component soliton cannot form for $\mratio=1:3$
(Figs. \ref{fig:1to3_SP}--\ref{fig:1to3_SMRD}).
}
\label{fig:EPMPj}
\end{figure*}

\begin{figure}
\centering
\includegraphics[width=\columnwidth]{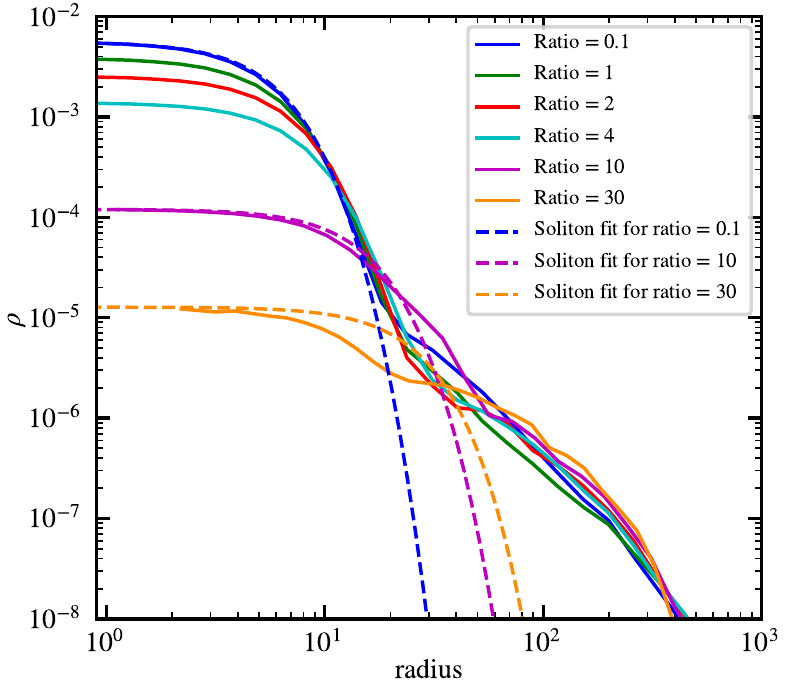}
\caption{
Density profiles of the six systems in \fref{fig:EPMPj} (solid lines).
The dashed lines represent the fitted analytical soliton solutions
considering the simple harmonic external potential for comparison.
The larger the energy ratio, the lower the peak density and mass of the
central object after merger. Especially, the density profiles
start to deviate from the soliton solution when the energy ratio exceeds
$\sim 10$, suggesting that a soliton is difficult to form in a hot environment
under a deep potential, consistent with our cosmological simulations with
$\mratio=1:3$ (Figs. \ref{fig:1to3_SP}--\ref{fig:1to3_SMRD}).
}
\label{fig:EPMPf}
\end{figure}

\begin{figure}
\centering
\includegraphics[width=\columnwidth]{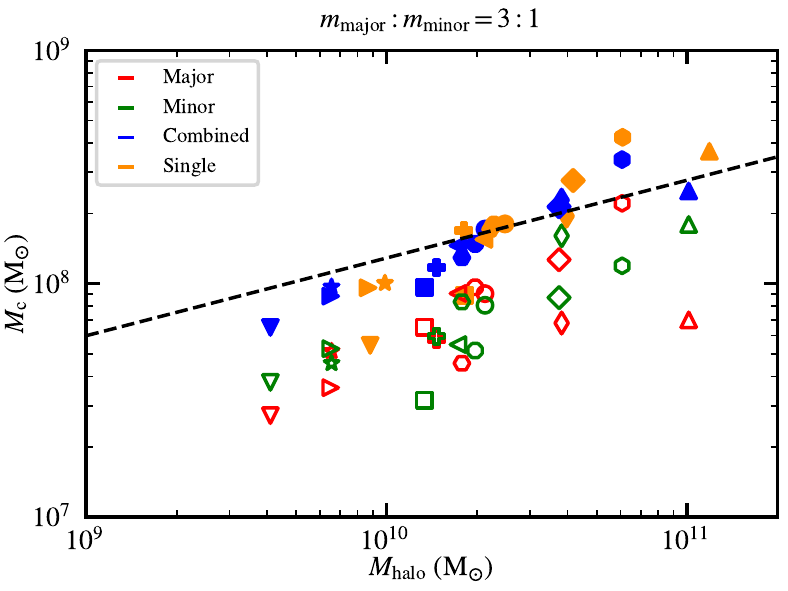}
\caption{
Core-halo mass relation at $z=0$ for $\mratio=3:1$.
The same halo of different components is represented by the same symbol with different colors.
Even considering the scatter in the data points,
both the major-component (red) and minor-component (green) soliton masses
are clearly lower than their single-component counterpart (orange),
consistent with \fref{fig:3to1_P}.
However, the \emph{combined} soliton mass of the two components (blue)
is generally similar to the single-component soliton,
suggesting that the combined mass in a $\mratio=3:1$ system follows
the original single-component core-halo mass relation,
$M_{\rm c} \propto M_{\rm halo}^{1/3}$ (dashed line).
}
\label{fig:3to1_CHR}
\end{figure}

\subsection{Two-component soliton profiles}
\label{subsec:two_soliton}

We want to obtain the minor-component soliton profile under the influence of
a major-component soliton. To this end, we assume the major component still
follows the original soliton solution \eref{eq:soliton}, regard that
as a static external potential, and compute the corresponding minor-component
soliton solution by solving the time-independent, single-component
Schr\"{o}dinger-Poisson equation with spherical symmetry using the 4th order
Runge-Kutta method. Additionally, since both components live at the base
of the same halo potential, their velocity dispersions should be similar so
the length scales should be inversely proportional to $m$.
Therefore, we focus on the soliton solutions with a core radius ratio
$\rcM:\rcm \sim 1:3~(3:1)$ for $\mratio=3:1~(1:3)$.

\fref{fig:ATCSP} shows the resulting soliton density profiles.
For $\mratio=3:1$, we show the minor-component solitons with a peak density ratio
$\rho_{\rm c, major}$/$\rho_{\rm c, minor} = 1, 4, 16, 64, 256$,
corresponding respectively to a core radius ratio
$\rcM/\rcm = 0.64, 0.51, 0.44, 0.41, 0.39$
and a core mass ratio
$\McM/\Mcm = 0.27, 0.54, 1.38, 4.39, 16.10$.
With a lower peak density, the minor-component soliton radius increases
but the mass decreases, where $\rcM/\rcm$ is typically
lower than $0.5$ and approaching $\sim 0.4$ while $\McM/\Mcm$
can be greater or less than unity across a wide range of peak densities.
These results are consistent with our cosmological simulations,
where $\rcM/\rcm \sim 0.3\textrm{--}0.5$ (\fref{fig:3to1_RD})
and $\McM/\Mcm \sim 0.5\textrm{--}2.0$ (\fref{fig:3to1_SMRD}).

For $\mratio=1:3$, fixing $\rcM:\rcm = 3:1$ leads to
$\rho_{\rm c, major}$/$\rho_{\rm c, minor} = 1/7.3$ and
$\McM/\Mcm = 3.7$.
So the minor-component soliton appears as a small density clump
on the top of the major-component soliton (which inspires the
toy model presented in the next subsection).
The mass ratio will increase further with a larger
$\rho_{\rm c, major}$/$\rho_{\rm c, minor}$.
For example, for $\rho_{\rm c, major}$/$\rho_{\rm c, minor} = 1/3$,
$\McM/\Mcm$ increases to $5.5$.
In comparison, in our cosmological simulations the minor-component soliton
(i) has a peak density generally lower than $7.3\rho_{\rm c, major}$ during
the violent relaxation phase ($5 \gtrsim z \gtrsim 3$ in \fref{fig:1to3_PDE})
and (ii) does not re-emerge once the major-component soliton is stabilized
($z \lesssim 3$). It suggests that a minor-component soliton with $\mratio=1:3$
is unstable and difficult to form in a \emph{hot} environment
caused by the deep gravitational potential of the major-component soliton.
We discuss it further in the next subsection.

\subsection{Toy model for minor-component soliton formation}
\label{subsec:toy_model}

Here we propose a toy model to investigate why a minor-component soliton
cannot re-emerge with the presence of a stable major-component soliton for
$\mratio=1:3$ (Figs. \ref{fig:1to3_SP}--\ref{fig:1to3_SMRD}).
We let ten minor-component solitons merge in a static simple harmonic potential,
$\Phi(r) \propto r^2$, mimicking a major-component soliton.
All solitons are initially at rest and randomly distributed.
We experiment with different depths of the simple harmonic potential,
using the energy ratio between the simple harmonic potential energy and
the self-gravitational potential energy of all solitons in the initial
conditions as an indicator, and compare the final relaxed configurations
after merger. The simulations are dimensionless.

\fref{fig:EPMPj} and \fref{fig:EPMPf} show respectively the projected density
and density profiles of the final states with different initial energy ratios.
When the simple harmonic potential is deeper, the environment is hotter,
and the central object formed after merger becomes more diffusive and less
prominent. As the energy ratio exceeds $\sim 10$, the density profiles
start to deviate from the soliton solution and the soliton does not form during
the entire merging and relaxation process.
It indicates that it is difficult to form a soliton in a hot environment under
a deep potential, consistent with our cosmological simulations.
For $\mratio=1:3$, the gravitational potential near the halo centre is
dominated by the massive major-component soliton, which serves as an
external hot environment that hinders the formation of a minor-component soliton.
Here the \emph{hot} environment simply means the gravitational potential is deep.

The above finding has some intriguing implications. For example, \citet{Luu2020}
suggest that the nuclear star cluster of the Milky Way can be
explained by nested solitons from two $\psiDM$ components,
with $\mM \sim 10^{-22}\eV$ and $\mm \sim 10^{-20}\eV$.
However, if our results could be extrapolated to a higher particle mass ratio
and different dark matter density fraction, it would imply that such a
higher-particle-mass minor-component soliton cannot form.
Another important question concerns whether a \emph{single}-component
soliton can form under the deep gravitational potential of a baryonic bulge
with mass $\sim 10^{10} \Msun$, an order of magnitude more massive
than a typical soliton in a Milky Way-sized halo. If it cannot form,
the conventional core-halo relation \citep[e.g.][]{Schive2014PhRvL}
must take into account this diversity.

\subsection{Two-component core-halo relation}
\label{subsec:core_halo}

A natural follow-up question is whether there is a relation between the
two-component solitons and their host haloes.
For $\mratio=1:3$, the minor-component core is negligible (\fref{fig:1to3_SMRD}),
while the major-component soliton is similar to its single-component counterpart
(\fref{fig:1to3_P}). So the major component (or the sum
of both components since the minor-component core mass is negligible anyway)
should follow the same single-component core-halo mass relation,
$M_{\rm c} \propto M_{\rm halo}^{1/3}$ \citep{Schive2014PhRvL}.
But what about $\mratio=3:1$, for which the major- and minor-component
soliton masses are comparable (\fref{fig:3to1_SMRD})?

\fref{fig:3to1_CHR} shows $M_{\rm c}$ vs. $M_{\rm halo}$ for $\mratio=3:1$
and the single-component case.
Even considering the scatter in the data points,
both the major- and minor-component soliton masses are clearly lower than their
single-component counterpart, consistent with \fref{fig:3to1_P}.
However, to our surprise, the \emph{combined} soliton mass of the two
components, $M_{\rm c, combined} = \McM + \Mcm$,
is generally similar to the single-component soliton mass.
It suggests that there is a more general core-halo relation where the soliton mass
in the original single-component core-halo relation can be replaced by
the combined soliton mass of two components (for both $\mratio=3:1$ and $1:3$).
This is also a new finding to show the soliton thermodynamics has a serious problem
when extended to multi-component cases.
The soliton mass is expected to be an extensive quantity
in soliton thermodynamics since it is mainly controlled by
the virial temperature in the halo
(even though there may be small scatter in the relation).
If we have N components, the total soliton mass will be $\sum_{{\rm n}=1}^{\rm N} M_{\rm c,n}$,
which is much bigger than the single-component soliton mass for a given halo mass.
This is however not the case in our simulation result.
On the contrary, our result seems to indicate the soliton mass is an
intensive quantity in the thermodynamic sense.
This problem and the soliton mass ratio problem (\fref{fig:3to1_SMRD})
both indicate the original single-component soliton thermodynamics fails.

Furthermore, we propose that the soliton mass of individual component
($\McM$ or $\Mcm$) is proportional to its initial
mass density fraction ($75$ and $25$ per cent for the major and minor components,
respectively) and inversely proportional to the dark matter particle mass
($\mM$ or $\mm$). This conjecture leads to $\McM:\Mcm=1:1$
for $\mratio=3:1$ and $\McM:\Mcm=10:1$ for $\mratio=1:3$,
qualitatively consistent with our cosmological simulations.

However, we emphasize that, unlike the single-component case, the above
two-component core-halo relation is purely empirical and requires further
investigation and justification in the future.

\subsection{Minor-component power spectrum}
\label{subsec:minor_power}

The addition of a minor component can reshape the total $\psiDM$ density
power spectrum (\fref{fig:PS}), which may help alleviate some existing
tension between theories and observations. For example,
the power spectrum of $\mratio=1:3$ at $z=10$ is noticeably higher than the
single-component case at high $k$. This enhanced small-structure power in
filaments at high redshifts may diminish the tension
between Lyman-$\alpha$ forest observations and $\psiDM$ predictions
\citep[e.g.][]{PhysRevLett.119.031302},
although a variation of $\psiDM$, the extreme-axion model,
is also able to alleviate the inconsistency
 \citep{Leong2019MNRAS,Zhang2017PhRvDa,Zhang2017PhRvDb}.
On the other hand, the power spectrum of $\mratio=3:1$ grows slower than the
single-component case at high $k$. If our results are extrapolated to a
higher particle mass ratio case (e.g. $\mratio=15:1$), it may provide a
possible solution to the $\sigma_8$ tension \citep[e.g.][]{Abdalla2022}
that the observed matter fluctuations smoothed over $8 \Mpch$ are smaller
than the $\Lambda$CDM prediction.

\section{Conclusion}
\label{sec:concl}

In this work, we explore a two-component $\psiDM$ scenario via cosmological
simulations. The two components are described by two separate scalar fields
coupled only by gravity, and their evolution is governed by the coupled
Schr\"{o}dinger-Poisson equations (\ref{eq:schrodinger1})--(\ref{eq:poisson}).
By utilizing the adaptive mesh refinement code $\gamer$ to achieve sufficient
high resolution (\fref{fig:3to1_HS}), we focus on the non-linear structures
of two-component haloes and solitons in a $1.4 \Mpch$ comoving box.
We assume the total dark matter mass contains $75$ per cent major component
and $25$ per cent minor component, fix the major-component particle mass to
$\mM=1\times10^{-22}\eV$, and explore two different minor-component
particle masses, $\mratio=3:1$ and $1:3$.
Our main results are summarized as follows:

\begin{itemize}
\item On large scales, the spatial distributions of filaments and massive haloes
      are very similar between the two- and single-component models
      (\fref{fig:3to1_GP} and \fref{fig:1to3_GP}).

\item For both $\mratio=3:1$ and $1:3$, haloes are composed of both
      components, and each of which exhibits distinct wave features
      (e.g. density granulation) with a characteristic length scale
      inversely proportional to the particle mass $m$
      (\fref{fig:3to1_HP} and \fref{fig:1to3_HP}). The outskirts of the
      two-component halo density profile closely follow the single-component
      counterpart (Figs. \ref{fig:3to1_P} and \ref{fig:1to3_P}).

\item In $\mratio=3:1$ haloes, major- and minor-component solitons
      coexist and are roughly concentric
      (Figs. \ref{fig:3to1_SP} and \ref{fig:3to1_RD}).
      The major-component soliton follows the original soliton profile
      (\eref{eq:soliton}) while the minor-component soliton follows
      a compressed soliton profile (\fref{fig:ATCSP}); both components
      contribute comparably to the soliton total density profile.
      The soliton peak density is significantly lower than the
      single-component counterpart, which, together with the presence of
      an extended minor-component soliton, leads to a much smoother
      soliton-to-halo transition (\fref{fig:3to1_P}).
      As a result, the bump in the rotation velocity stemming from a massive
      and compact single-component soliton becomes much less prominent in a
      $\mratio=3:1$ halo (\fref{fig:RC}), which can alleviate the tension
      with some observations \citep[e.g.][]{Bar2022}.

\item In $\mratio=3:1$ haloes, the masses of the major- and minor-component
      solitons are comparable, with ratios ranging mostly between
      $0.5 \textrm{--} 2.0$ and independent of halo mass
      (\fref{fig:3to1_SMRD}). This result is consistent with the two-component
      soliton solutions (\fref{fig:ATCSP}).
      The combined mass of the major- and minor-component solitons is found
      to follow the original single-component core-halo mass relation,
      $M_{\rm c, combined} = \McM + \Mcm \propto M_{\rm halo}^{1/3}$
      (\fref{fig:3to1_CHR}).
      These results indicate the single-component soliton thermodynamic expectation
      fails to extend to the multi-component cases.

\item In $\mratio=1:3$ haloes, the minor-component soliton cannot form
      once the major-component soliton is stabilized
      (Figs. \ref{fig:1to3_SP}--\ref{fig:1to3_SMRD}).
      This surprising result can be explained by the toy model in
      \sref{subsec:toy_model}, showing that a soliton cannot re-emerge after
      falling in a deep simple harmonic potential
      (Figs. \ref{fig:EPMPj} and \ref{fig:EPMPf}).
      It suggests that it is difficult to form a soliton in a hot environment
      associated with a deep potential.
      The total density profile in $\mratio=1:3$, for both halo and soliton,
      is thus dominated by the major component and closely follows the
      single-component profile (\fref{fig:1to3_P}).
\end{itemize}

The above findings provide a set of predictions that sheds light on a
multi-component $\psiDM$ model. For future work, we can explore different
particle mass ratios and different total mass ratios.
What is the critical ratio to show a significant difference compared with
the single-component scenario?
Would a minor component with a much larger particle mass (e.g. $\mratio=1:10$)
collapse earlier and form stable solitons?
What happens if there are more than two components or even a continuous
distribution of different particle masses?
Last but not least, we want to explore whether a massive bulge can provide a
hot environment to disrupt the soliton even in the single-component
scenario.

\section*{Acknowledgements}
\label{sec:acknowledgements}

We thank to National Center for High-performance Computing (NCHC)
for providing computational and storage resources.
H. S. acknowledges funding support from the Jade Mountain Young Scholar Award
No. NTU-111V1201-5, sponsored by the Ministry of Education, Taiwan.
This research is partially supported by the
National Science and Technology Council (NSTC) of Taiwan under
Grants No. NSTC 111-2628-M-002-005-MY4, No. NSTC 108-2112-M-002-023-MY3,
and No. NSTC-110-2112-M-002-018,
and the NTU Academic Research-Career Development Project under
Grant No. NTU-CDP-111L7779.

\section*{Data Availability}
\label{sec:data_availability}

The data underlying this article will be shared
on reasonable request to the corresponding author.

\bibliographystyle{mnras}
\bibliography{ref}

\appendix

\section{Code verification}
\label{app:code_verify}

To verify the accuracy of our cosmological two-component $\psiDM$ code,
we conduct the following two numerical tests:
two-component Jeans instability test (\sref{sub:TwoJeansInstability}) and
two-component solitons test (\sref{sub:TwoSolitonsConcentric}).

\subsection{Two-component Jeans instability test}
\label{sub:TwoJeansInstability}

We review the single-component $\psiDM$ Jeans instability in
\sref{subsub:SingleJeans}, generalize it to two components in
\sref{subsub:TwoJeans} and conduct numerical experiments in
\sref{subsub:TwoJeansTest}.

\begin{figure*}
\centering
\includegraphics[width=\textwidth]{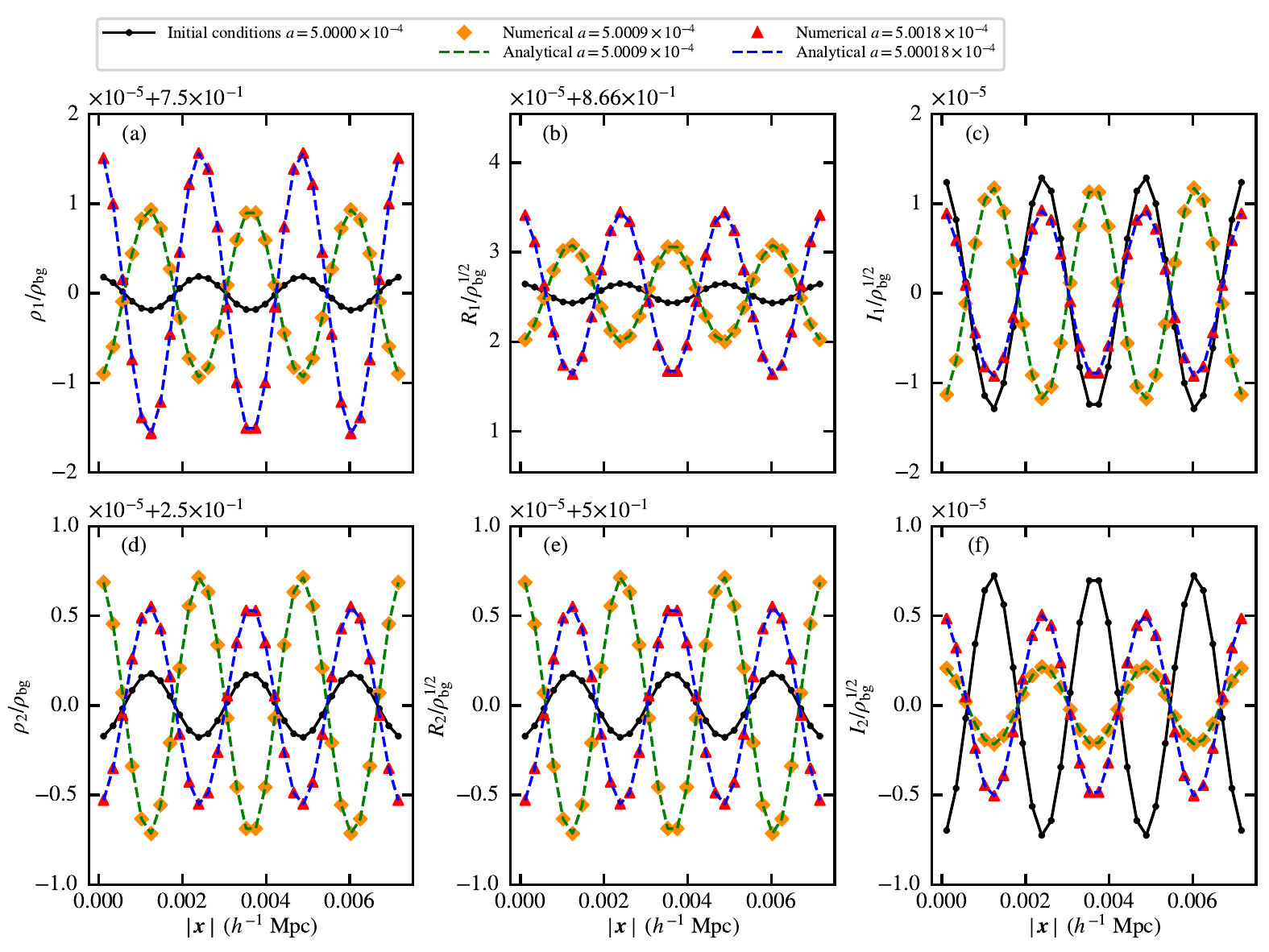}
\caption{
Two-component Jeans instability test: stable case.
Panels (a)--(c) show respectively the density, real part, imaginary part
of component 1 and panels (d)--(f) show the results of component 2.
Diamonds and triangles show the numerical results at two different redshifts.
Dashed lines represent the corresponding analytical solutions
(Eqs. (\ref{eq:TwoJeansSolutionStable}), (\ref{eq:TwoJeansReal1})--(\ref{eq:TwoJeansDens2})).
Black solid lines with dots give the initial conditions for reference.
The simulation set-up is described in \sref{subsub:TwoJeansTest}.
Perturbations are oscillating in this case.
The simulation end time $a=0.00050018$ corresponds to $\sim 1.1$ oscillation
periods for component 1 and more than $3$ periods for component 2.
The numerical results agree well with the analytical solutions.
}
\label{fig:TwoJeansStable}
\end{figure*}

\begin{figure*}
\centering
\includegraphics[width=\textwidth]{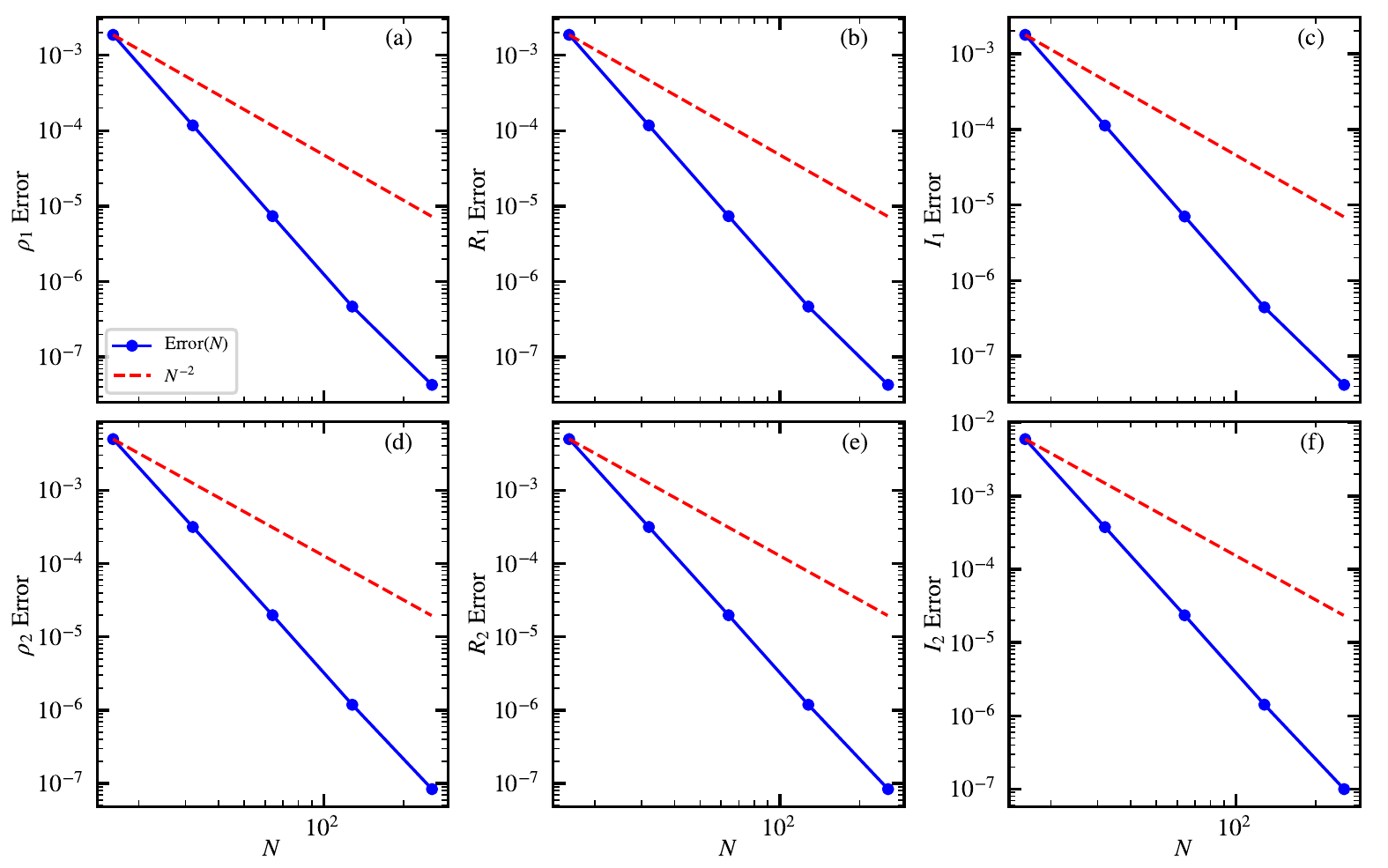}
\caption{
Error convergence in the two-component Jeans instability test: stable case.
Panels (a)--(c) show respectively the errors of density, real part,
imaginary part of component 1 and panels (d)--(f) show the errors of
component 2.
Blue solid lines with dots show the simulation results,
which are better than second-order accuracy (red dashed lines)
in this test.
}
\label{fig:TwoJeansStableL1Error}
\end{figure*}

\begin{figure*}
\centering
\includegraphics[width=\textwidth]{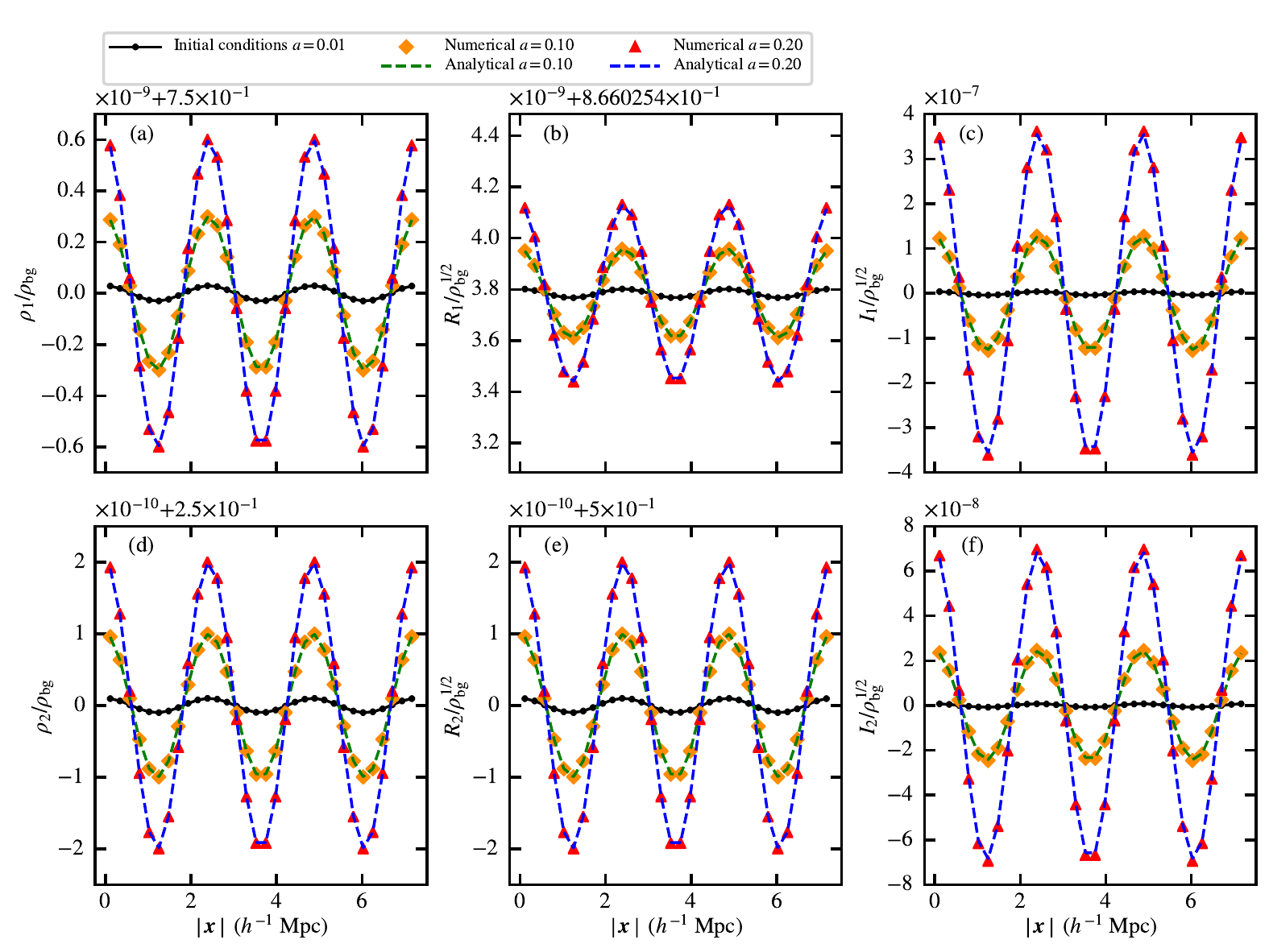}
\caption{
Similar to \fref{fig:TwoJeansStable} but for the unstable case.
The analytical solutions are given in Eqs.
(\ref{eq:TwoJeansSolutionUnstable})--(\ref{eq:TwoJeansDens2}).
Perturbations are unstable and growing linearly with $a$.
The simulation end time $a=0.2$ corresponds to when
the density perturbation amplitude has grown by a factor of 20.
The numerical results match well with the analytical solutions.
}
\label{fig:TwoJeansUnstable}
\end{figure*}

\begin{figure*}
\centering
\includegraphics[width=\textwidth]{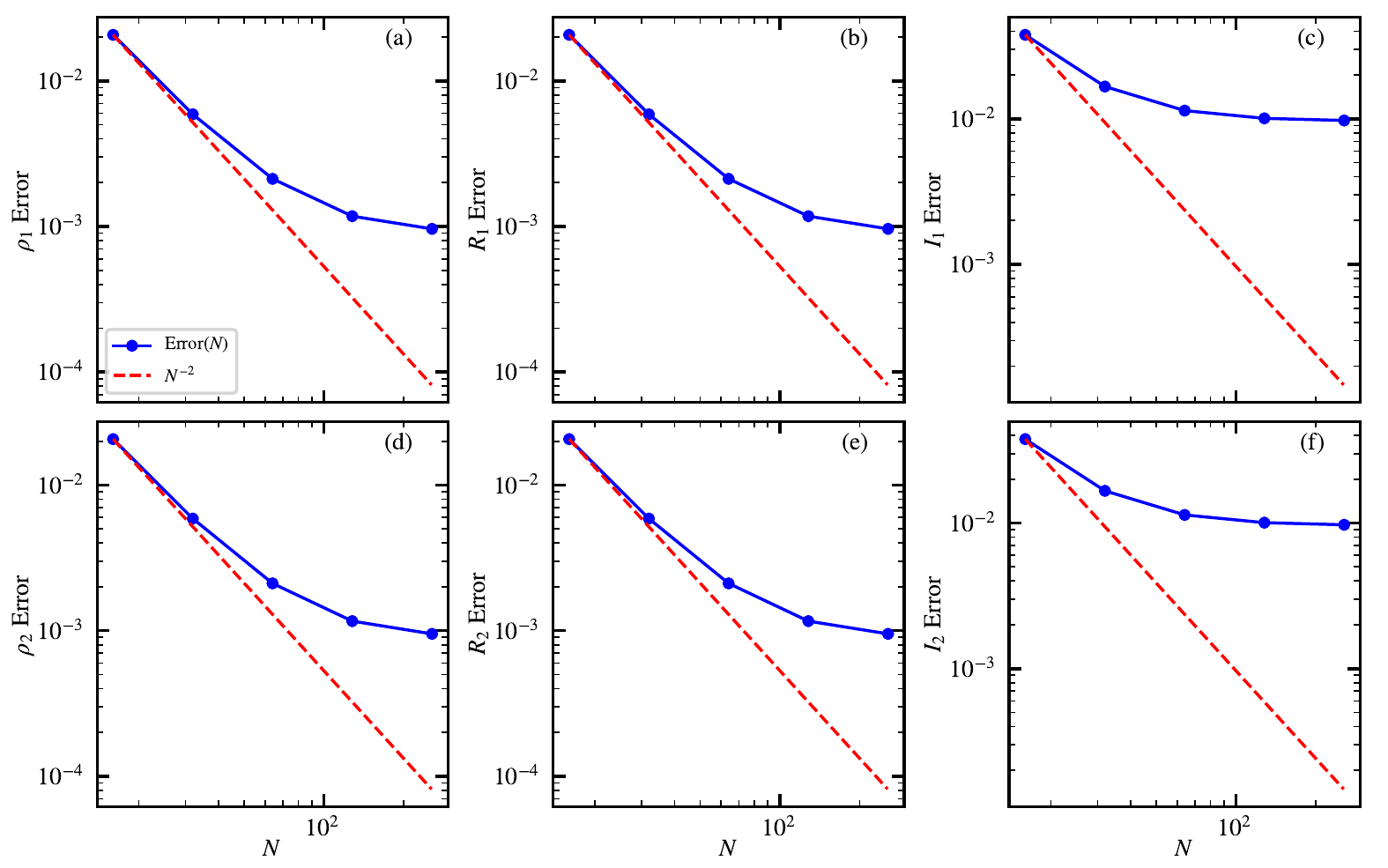}
\caption{
Similar to \fref{fig:TwoJeansStableL1Error} but for the unstable case.
The error convergence is roughly second-order at lower resolution
for the densities and real parts of both components.
But at higher resolution, especially for the imaginary parts,
errors converge much slower than second-order,
possibly due to linearization errors or round-off errors.
}
\label{fig:TwoJeansUnstableL1Error}
\end{figure*}

\subsubsection{Single-component Jeans instability}
\label{subsub:SingleJeans}

Assume the background density of the universe is $\rho_{\rm bg}$.
The overdensity is defined as
\be
\label{eq:SingleJeansDeltaDef}
\delta(\vect{x})=\frac{\Delta\rho(\vect{x})}{\rho_{\rm bg}},
\ee
where $\Delta\rho(\vect{x})$ is the density perturbation with the total
density given by $\rho(\vect{x})=\rho_{\rm bg}+\Delta\rho(\vect{x})$.
After Fourier transformation, $\delta(\vect{x})$ can be decomposed
into different $k$ modes. For simplicity, we only keep the cosine terms,
\be
\label{eq:SingleJeansDeltaKDef}
\delta(\vect{x})=\sum_{k}\delta_{k}\cos(\vect{k}\cdot\vect{x}).
\ee
By introducing
\be
\label{eq:SingleJeansYDef}
y=\frac{\hbar k^2}{mH_0\sqrt{a}}=\sqrt{6}\frac{k^2}{k_{\rm J}^2},
\ee
where $a$ is the scale factor of cosmic expansion, $m$ is the
$\psiDM$ particle mass, $H_0$ is the current Hubble parameter, and
\be
\label{eq:SingleJeansKJDef}
k_{\rm J}=\Big(\frac{6aH_0^2m^2}{\hbar^2}\Big)^{1/4}
\ee
is the Jeans wavenumber,
the linearized equation for the evolution of wave dark matter
overdensity can be written as \citep{Woo2009ApJ}
\be
\label{eq:SingleJeansEquation}
y^2\frac{{\rm d}^2}{{\rm d}y^2}\delta_k+(y^2-6)\delta_k=0.
\ee
The general solution to \eref{eq:SingleJeansEquation} is
\be
\label{eq:SingleJeansSolution}
\begin{aligned}
\delta_k=&c_{+}\frac{3\cos(y)+3y\sin(y)-y^2\cos(y)}{y^2}+\\
&c_{-}\frac{3\sin(y)-3y\cos(y)-y^2\sin(y)}{y^2},
\end{aligned}
\ee
where $c_+$ and $c_-$ are the coefficients of the growing and
decaying modes, respectively.
Given a fixed $k$, $y$ decreases when $a$ increases.

The critical wavenumber $k=k_{\rm J}$ corresponds to $y=y_{\rm J}=\sqrt{6}$.
On small scales with $k>k_{\rm J}$ ($y>y_{\rm J}$), perturbations are stable
and oscillating since quantum pressure balances gravity.
In the limit $y\gg1$, the solution \eref{eq:SingleJeansSolution} reduces to
\be
\label{eq:SingleJeansSolutionStable}
\delta_k=-c_{+}\cos(y)-c_{-}\sin(y).
\ee
In contrast, on large scales with $k<k_{\rm J}$ ($y<y_{\rm J}$),
perturbations are unstable and can collapse since gravity dominates over
quantum pressure, known as the Jeans instability.
For $y\ll1$, the solution \eref{eq:SingleJeansSolution} reduces to
\be
\label{eq:SingleJeansSolutionUnstable}
\delta_k=c_{+}\frac{3}{y^2}+c_{-}\frac{y^3}{15},
\ee
where the growing mode evolves as $\delta_k \propto y^{-2} \propto a$,
same as CDM.

Assume the wave function takes the form
$\psi = R_{\rm bg} + \Delta R + i\Delta I$,
where $R$ and $I$ are the real and imaginary parts, respectively.
The corresponding linearized solution of the wave function perturbations
$\Delta R$ and $\Delta I$ can be derived in a similar way \citep{Woo2009ApJ}.
The perturbation of real part $R_k$ follows the same equation
as the density perturbation \eref{eq:SingleJeansEquation}.
The perturbation of imaginary part satisfies $I_k=-{\rm d}R_k/{\rm d}y$.
For a single-mode perturbation,
the real part of the wave function is
\be
\label{eq:SingleJeansReal}
R=R_{\rm bg}+R_k\cos(\vect{k}\cdot\vect{x})
=\sqrt{\rho_{\rm bg}}
+\frac{1}{2}\sqrt{\rho_{\rm bg}}\delta_{k}\cos(\vect{k}\cdot\vect{x}),
\ee
the imaginary part is
\be
\label{eq:SingleJeansImag}
I=I_k\cos(\vect{k}\cdot\vect{x})
=-\frac{{\rm d}R_k}{{\rm d}y}\cos(\vect{k}\cdot\vect{x})
\ee
and the density is
\be
\label{eq:SingleJeansDens}
\begin{aligned}
\rho &= R^2+I^2\\
&=\rho_{\rm bg}+\rho_{\rm bg}\delta_{k}\cos(\vect{k}\cdot\vect{x})
+I_k^2\cos^2(\vect{k}\cdot\vect{x})
+\mathcal{O}(\delta_{k}^2).
\end{aligned}
\ee
Note that the linearization assumes $I_k^2 \ll 2R_k \ll R_{\rm bg}$.

\subsubsection{Two-component Jeans instability}
\label{subsub:TwoJeans}

Here we extend the original single-component Jeans instability to two
components. The background density fractions of components 1 and 2 are
given respectively by
$f_1=\rho_{\rm bg,1}/\rho_{\rm bg}$ and
$f_2=\rho_{\rm bg,2}/\rho_{\rm bg}$,
where $0\leq f_1\leq 1$, $0\leq f_2\leq 1$ and $f_1+f_2=1$.
The density distribution of each component can be written as
\be
\left\{
\label{eq:TwoJeansRhoDef}
\begin{array}{lr}
\rho_1(\vect{x})=\rho_{\rm bg,1}+\Delta\rho_1(\vect{x})\\
\rho_2(\vect{x})=\rho_{\rm bg,2}+\Delta\rho_2(\vect{x}),
\end{array}
\right.
\ee
where $\Delta\rho_1(\vect{x})$ and $\Delta\rho_2(\vect{x})$
are the perturbations of individual components,
corresponding to overdensities
\be
\label{eq:TwoJeansDeltaDef}
\left\{
\begin{array}{lr}
\delta_1(\vect{x})=\frac{\Delta\rho_1(\vect{x})}{\rho_{\rm bg,1}}\\
\delta_2(\vect{x})=\frac{\Delta\rho_2(\vect{x})}{\rho_{\rm bg,2}}.
\end{array}
\right.
\ee
The total overdensity is then given by
\be
\label{eq:TwoJeansDeltaTot}
\delta_{\rm tot}
=\frac{\Delta\rho_1(\vect{x})+\Delta\rho_2(\vect{x})}{\rho_{\rm bg}}
=f_1\delta_1+f_2\delta_2.
\ee

After Fourier transformation, individual overdensity can be decomposed
into different $k$ modes
\be
\label{eq:TwoJeansDeltaKDef}
\left\{
\begin{array}{lr}
\begin{aligned}
\delta_1=\sum_{k}\delta_{k,1}\cos(\vect{k}\cdot\vect{x})\\
\delta_2=\sum_{k}\delta_{k,2}\cos(\vect{k}\cdot\vect{x}),
\end{aligned}
\end{array}
\right.
\ee
where for simplicity we only keep the cosine terms
and assume the two-component perturbations are in phase.
The Jeans wavenumber of each component is
\be
\label{eq:TwoJeansKJDef}
\left\{
\begin{array}{lr}
k_{\rm J,1}=\Big(\frac{6aH_0^2m_1^2}{\hbar^2}\Big)^{1/4}\\
k_{\rm J,2}=\Big(\frac{6aH_0^2m_2^2}{\hbar^2}\Big)^{1/4},
\end{array}
\right.
\ee
where $m_1$ and $m_2$ are the respective particle masses.
Similar to \eref{eq:SingleJeansYDef}, we define
$y_1$ and $y_2$ as
\be
\label{eq:TwoJeansYDef}
\left\{
\begin{array}{lr}
y_1=\frac{\hbar k^2}{m_1H_0\sqrt{a}}
=\sqrt{6}\frac{k^2}{k_{\rm J,1}^2}=\frac{m}{m_1}y\\
y_2=\frac{\hbar k^2}{m_2H_0\sqrt{a}}
=\sqrt{6}\frac{k^2}{k_{\rm J,2}^2}=\frac{m}{m_2}y.
\end{array}
\right.
\ee
The two-component linearized equations coupled through a common gravity read
\be
\label{eq:TwoJeansEquation}
\left\{
\begin{array}{lr}
y^2\frac{{\rm d}^2}{{\rm d}y^2}\delta_{k,1}
+[(\frac{m}{m_1}y)^2-6f_1]\delta_{k,1}-6f_2\delta_{k,2}=0\\
y^2\frac{{\rm d}^2}{{\rm d}y^2}\delta_{k,2}
+[(\frac{m}{m_2}y)^2-6f_2]\delta_{k,2}-6f_1\delta_{k,1}=0.
\end{array}
\right.
\ee
Here we use the same $y$ in the single-component case
(\eref{eq:SingleJeansYDef}) as the independent variable
and keep the single-component $m$ as a reference particle mass
in order to write the equations in a more symmetric form.

We do not find an exact solution to \eref{eq:TwoJeansEquation}.
However, an approximate solution can be obtained under the
small-scale or large-scale limit.
In the small-scale limit with $\frac{m}{m_1}y\gg1$ and $\frac{m}{m_2}y\gg1$,
the two components become decoupled and \eref{eq:TwoJeansEquation} reduces to
\be
\label{eq:TwoJeansEquationLargeY}
\left\{
\begin{array}{lr}
y^2\frac{{\rm d}^2}{{\rm d}y^2}\delta_{k,1}
+(\frac{m}{m_1}y)^2\delta_{k,1}=0\\
y^2\frac{{\rm d}^2}{{\rm d}y^2}\delta_{k,2}
+(\frac{m}{m_2}y)^2\delta_{k,2}=0.
\end{array}
\right.
\ee
The solutions to \eref{eq:TwoJeansEquation} are
\be\
\label{eq:TwoJeansSolutionStable}
\left\{
\begin{array}{lr}
\delta_{k,1}=c_1\cos(\frac{m}{m_1}y)+c_2\sin(\frac{m}{m_1}y)\\
\delta_{k,2}=c_3\cos(\frac{m}{m_2}y)+c_4\sin(\frac{m}{m_2}y),
\end{array}
\right.
\ee
where $c_1$, $c_2$, $c_3$ and $c_4$ are coefficients determined by the
initial conditions.
In this limit, both components are stable and oscillating with
their own frequency, similar to \eref{eq:SingleJeansSolutionStable}.

In the large-scale limit with $\frac{m}{m_1}y\ll1$ and $\frac{m}{m_2}y\ll1$,
\eref{eq:TwoJeansEquation} reduces to
\be
\label{eq:TwoJeansEquationSmallY}
\left\{
\begin{array}{lr}
y^2\frac{{\rm d}^2}{{\rm d}y^2}\delta_{k,1}
-6f_1\delta_{k,1}-6f_2\delta_{k,2}=0\\
y^2\frac{{\rm d}^2}{{\rm d}y^2}\delta_{k,2}
-6f_2\delta_{k,2}-6f_1\delta_{k,1}=0.
\end{array}
\right.
\ee
The solutions to \eref{eq:TwoJeansEquationSmallY} are
\be
\label{eq:TwoJeansSolutionUnstable}
\left\{
\begin{array}{lr}
\delta_{k,1}=c_5\frac{1}{y^2}-c_6f_2-c_7f_2y+c_8y^3\\
\delta_{k,2}=c_5\frac{1}{y^2}+c_6f_1+c_7f_1y+c_8y^3,
\end{array}
\right.
\ee
where $c_5$, $c_6$, $c_7$ and $c_8$ are coefficients
determined by the initial conditions.
There are two extra terms in this solution
compared to \eref{eq:SingleJeansSolutionUnstable}.
In this limit, both components are unstable.

Similar to the single-component case Eqs.
(\ref{eq:SingleJeansReal})--(\ref{eq:SingleJeansDens}),
we can express the linearized solutions
of the two-component wave functions explicitly as
the component-1 real part
\be
\label{eq:TwoJeansReal1}
R_1=R_{\rm bg,1}+R_{k,1}\cos(\vect{k}\cdot\vect{x})
=\sqrt{f_1\rho_{\rm bg}}
+\frac{1}{2}\sqrt{f_1\rho_{\rm bg}}\delta_{k,1}\cos(\vect{k}\cdot\vect{x}),
\ee
the component-1 imaginary part
\be
\label{eq:TwoJeansImag1}
I_1=I_{k,1}\cos(\vect{k}\cdot\vect{x})
=-\frac{m_1}{m}\frac{{\rm d}R_{k,1}}{{\rm d}y}\cos(\vect{k}\cdot\vect{x}),
\ee
the component-1 density
\be
\begin{aligned}
\label{eq:TwoJeansDens1}
\rho_1 &= R_1^2+I_1^2\\
&=f_1\rho_{\rm bg}+f_1\rho_{\rm bg}\delta_{k,1}\cos(\vect{k}\cdot\vect{x})
+I_{k,1}^2\cos^2(\vect{k}\cdot\vect{x})
+\mathcal{O}(\delta_{k,1}^2),
\end{aligned}
\ee
the component-2 real part
\be
\label{eq:TwoJeansReal2}
R_2=R_{\rm bg,2}+R_{k,2}\cos(\vect{k}\cdot\vect{x})
=\sqrt{f_2\rho_{\rm bg}}
+\frac{1}{2}\sqrt{f_2\rho_{\rm bg}}\delta_{k,2}\cos(\vect{k}\cdot\vect{x}),
\ee
the component-2 imaginary part
\be
\label{eq:TwoJeansImag2}
I_2=I_{k,2}\cos(\vect{k}\cdot\vect{x})
=-\frac{m_2}{m}\frac{{\rm d}R_{k,2}}{{\rm d}y}\cos(\vect{k}\cdot\vect{x})
\ee
and the component-2 density
\be
\label{eq:TwoJeansDens2}
\begin{aligned}
\rho_2 &= R_2^2+I_2^2\\
&=f_2\rho_{\rm bg}+f_2\rho_{\rm bg}\delta_{k,2}\cos(\vect{k}\cdot\vect{x})
+I_{k,2}^2\cos^2(\vect{k}\cdot\vect{x})
+\mathcal{O}(\delta_{k,2}^2).
\end{aligned}
\ee
The linearization is valid when $I_{k,1}^2\ll2R_{k,1}\ll R_{\rm bg,1}$ and $I_{k,2}^2\ll2 R_{k,2}\ll R_{\rm bg,2}$.

\subsubsection{Numerical tests}
\label{subsub:TwoJeansTest}

We utilize the aforementioned two-component Jeans instability problem
to verify the cosmological code used in this work.
We test both the stable and unstable solutions, which share the following
set-up.
The matter density parameter is $\Omega_m = 1.0$
and the dimensionless Hubble parameter is $h=0.6732$.
We assume a single-mode perturbation with a wavevector
$\vect{k}=(k/\sqrt{3},k/\sqrt{3},k/\sqrt{3})$
along the diagonal of the computational domain.
The perturbation wavelength is $\lambda=L/\sqrt{3}$,
where $L$ is the comoving box size, such that there are three waves
along the diagonal.
The two components share the same wavevector and are in phase.
To mimic our cosmological simulations, we adopt
$m_1=1\times10^{-22}\eV$, $m_2=\frac{1}{3}\times10^{-22}\eV$,
$f_1=0.75$ and $f_2=0.25$.
The reference particle mass in \eref{eq:TwoJeansYDef}
is $m=1\times10^{-22}\eV$.

First, we test the stable case.
The analytical solutions are given by
Eqs. (\ref{eq:TwoJeansSolutionStable}), (\ref{eq:TwoJeansReal1})--(\ref{eq:TwoJeansDens2}),
where the coefficients are set to
$c_1=-3.0\times10^{-5}$, $c_2=0$, $c_3=-3.0\times10^{-5}$, $c_4=0$
and the perturbation amplitudes are
$R_{k,1}= 1.09\times10^{-6}\sqrt{\rho_{\rm bg}}$,
$I_{k,1}= 1.29\times10^{-5}\sqrt{\rho_{\rm bg}}$,
$R_{k,2}= -1.79\times10^{-6}\sqrt{\rho_{\rm bg}}$,
$I_{k,2}= -7.28\times10^{-6}\sqrt{\rho_{\rm bg}}$.
The comoving box size is $L=0.0042\Mpch$, corresponding to a
wavenumber $k=2591.14\MpchInv$.
We evolve the system from $a=0.0005$ to $0.00050018$ such that
component 1 oscillates for $\sim 1.1$ periods.
During this time span,
$y_1$ evolves from $3.8753\times10^{4}$ to $3.8746\times10^{4}$ and
$y_2$ evolves from $1.16259\times10^{5}$ to $1.16238\times10^{5}$.
The initial Jeans wavenumbers are $k_{\rm J,1}= 20.60\MpchInv$ and
$k_{\rm J,2}= 11.89\MpchInv$, satisfying
$k \gg k_{\rm J,1}$ and $k \gg k_{\rm J,2}$.
\fref{fig:TwoJeansStable} compares between the simulation results
with $N^3=32^3$ cells and analytical solutions,
demonstrating good agreement.
\fref{fig:TwoJeansStableL1Error} shows the error convergence
with different $N$.
Here we define the error as
\be
\label{eq:L1Error}
\mathrm{Error} = \frac{1}{N}\sum_{\rm i=1}^N
\left|\frac{\rm Numerical_i - \rm Analytical_i}
{\rm Analytical_i-\rm Analytical_{\rm bg}}\right|,
\ee
where `$\rm i$' is the cell index along the diagonal,
`$\rm Numerical_i$' is the simulation data,
`$\rm Analytical_i$' is the analytical solution and
`$\rm Analytical_{\rm bg}$' represents the background term.
It shows that the numerical accuracy is better than
second-order in this test.

Second, we test the unstable case.
The analytical solutions are given by
Eqs. (\ref{eq:TwoJeansSolutionUnstable})--(\ref{eq:TwoJeansDens2}),
where the coefficients are set to
$c_5=3.0\times10^{-15}$, $c_6=0$, $c_7=0$, $c_8=0$
to keep only the growing mode.
The initial perturbation amplitudes are
$R_{k,1}=1.73\times10^{-11}\sqrt{\rho_{\rm bg}}$,
$I_{k,1}=3.99\times10^{-9}\sqrt{\rho_{\rm bg}}$,
$R_{k,2}=9.99\times10^{-12}\sqrt{\rho_{\rm bg}}$,
$I_{k,2}=7.68\times10^{-10}\sqrt{\rho_{\rm bg}}$.
The comoving box size is $L=4.2\Mpch$, corresponding to a
wavenumber $k=2.59\MpchInv$.
We evolve the system from $a=0.01$ to $0.2$ such that
the density perturbation amplitudes of both components
will grow by a factor of 20.
During this time span,
$y_1$ evolves from $ 8.67\times10^{-3}$ to $1.94\times10^{-3}$ and
$y_2$ evolves from $ 2.60\times10^{-2}$ to $5.81\times10^{-3}$.
The initial Jeans wavenumbers are $k_{\rm J,1}= 43.56\MpchInv$ and
$k_{\rm J,2}= 25.15\MpchInv$, obeying
$k \ll k_{\rm J,1}$ and $k \ll k_{\rm J,2}$.
\fref{fig:TwoJeansUnstable} demonstrates good agreement between
the simulation results with $N^3=32^3$ cells and analytical solutions.
\fref{fig:TwoJeansUnstableL1Error} shows the error convergence
with different $N$.
It is roughly second-order at lower resolution
for the densities and real parts of both components.
But at higher resolution, especially for the imaginary parts,
errors converge much slower than second-order,
possibly due to linearization errors or round-off errors.

\subsection{Two-component solitons test}
\label{sub:TwoSolitonsConcentric}

\begin{figure*}
\centering
\includegraphics[width=\textwidth]{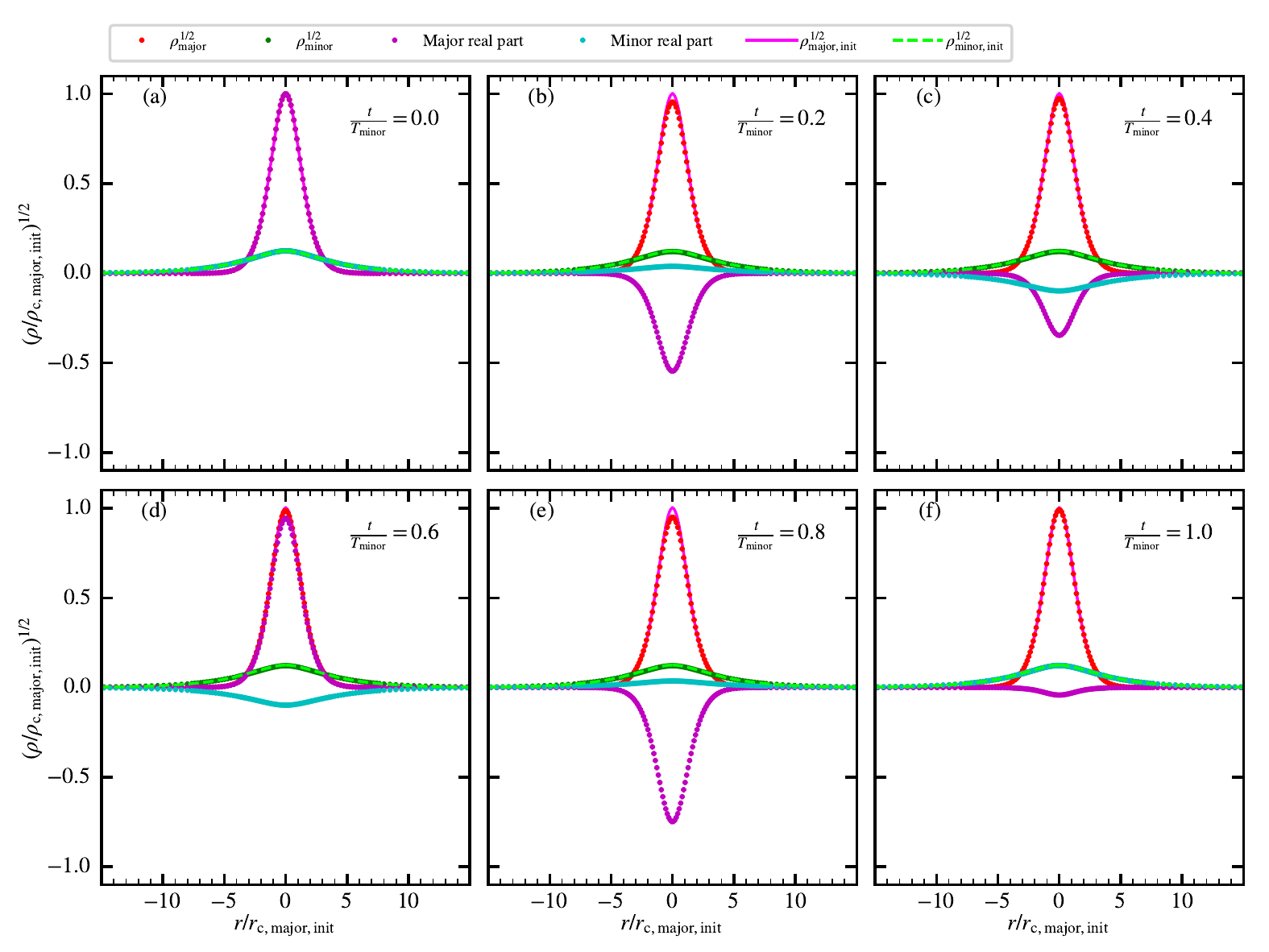}
\caption{
Evolution of two concentric solitons with distinct particle masses.
Panels (a)--(f) show the data along the diagonal of the box
at different time, where $T_{\rm minor}$ is the oscillation period of the
minor-component soliton wave function.
Red and green dots show respectively the square root of the major- and
minor-component density, and magenta solid and lime dashed lines show
the initial conditions for comparison.
Purple and cyan dots are the real parts of the major- and minor-component
wave functions, respectively.
It is expected that the soliton density is almost stationary but
the wave function is oscillating for both components.
At the simulation end time, the major-component (minor-component) soliton
wave function undergoes 8 (1) oscillations.
We find that both solitons are stable, confirming the accuracy of the
two-component $\psiDM$ code used in this work.
There is a $\sim 10$ per cent fluctuation in the peak density of
the major-component soliton, likely because the two-component solitons
are not constructed fully self-consistently (see text for details).
}
\label{fig:TwoSolitonsCocenter}
\end{figure*}

To further validate our two-component $\psiDM$ code in the fully non-linear
regime, we test the stability of two concentric solitons with distinct
particle masses via three-dimensional simulations.
We adopt $\mratio=3:1$ and $\rho_{\rm c, major}$/$\rho_{\rm c, minor} = 64$.
The compressed minor-component soliton is taken to be the same as the orange
line in \fref{fig:ATCSP} that considers the original (uncompressed)
major-component soliton (\eref{eq:soliton}) as an external potential.
To improve consistency, for the major-component soliton we further consider
the additional compression due to the external potential associated with the
minor-component soliton. Ideally, to be fully self-consistent, one should
repeat this iterative procedure until both soliton solutions converge.
However, it only has a small effect since the compact and massive major-component
soliton is insensitive to the minor-component potential.
So we do not apply this iterative method here.

We adopt a non-comoving box with a rigid wall boundary condition.
The simulation box has a size of $L \sim 72 \rcM$.
The base level resolution is $N=128$ and there are three refined AMR levels,
leading to $\sim 14$ cells for resolving $\rcM$.
\fref{fig:TwoSolitonsCocenter} shows the simulation results.
We confirm that both solitons remain stable after the major- and
minor-component soliton wave functions undergo 8 and 1 oscillations,
respectively.
There is a $\sim 10$ per cent fluctuation in the peak density of
the major-component soliton, likely because the two-component solitons
are not constructed fully self-consistently as mentioned above.
This test confirms the accuracy of (i) the compressed soliton solutions
and (ii) our two-component $\psiDM$ code.

\bsp
\label{lastpage}
\end{document}